\documentclass[usenatbib]{mn2e}
\usepackage{amsmath}

\usepackage{epsf}
\usepackage{bm}
\usepackage{graphicx}

\newcommand\bnabla{{\bmath\nabla}}
\newcommand\be{{\bmath e}}
\newcommand\bF{{\bmath F}}
\newcommand\bff{{\bmath f}}
\newcommand\bk{{\bmath k}}
\newcommand\bu{{\bmath u}}
\newcommand\bv{{\bmath v}}
\newcommand\bx{{\bmath x}}
\newcommand\half{{\textstyle\frac{1}{2}}}

\newcommand\rmd{\mathrm{d}}

\newcommand\rmi{\mathrm{i}}

\newcommand\rmt{\mathrm{t}}

\newcommand\f{\frac}
\newcommand\p{\partial}

\title[Tides and convection]
{On the interaction between tides and convection}

\author[Gordon I. Ogilvie and Geoffroy Lesur]
{Gordon I. Ogilvie$^{1}$ and Geoffroy Lesur$^{1,2}$\\
$^{1}$Department of Applied Mathematics and Theoretical Physics,
University of Cambridge, Centre for Mathematical Sciences,\\
Wilberforce Road, Cambridge CB3 0WA\\
$^{2}$UJF-Grenoble 1 / CNRS-INSU, Institut de Plan\'etologie et d'Astrophysique de Grenoble (IPAG) UMR 5274, \\Grenoble, F-38041, France
}

\begin{document}

\maketitle

\label{firstpage}
 
\begin{abstract}
  We study the interaction between tides and convection in
  astrophysical bodies by analysing the effect of a homogeneous
  oscillatory shear on a fluid flow.  {This model can be taken
    to represent the interaction between a large-scale periodic tidal
    deformation and a smaller-scale convective motion.}  We first
  consider analytically the limit in which the shear is of low
  amplitude and the oscillation period is short compared to the
  timescales of the {unperturbed} flow.  In this limit there is
  a viscoelastic response and we obtain expressions for the effective
  elastic modulus and viscosity coefficient.  The effective viscosity
  is inversely proportional to the square of the oscillation
  frequency, with a coefficient that can be positive, negative or zero
  depending on the properties of the unperturbed flow.  We also carry
  out direct numerical simulations of Boussinesq convection in an
  oscillatory shearing box and measure the time-dependent Reynolds
  stress.  The results indicate that the effective viscosity
  {of turbulent convection} falls rapidly as the oscillation
  frequency is increased, attaining small negative values in the cases
  we have examined, {although significant uncertainties remain
    because of the turbulent noise}.  We discuss the implications of
  this analysis for astrophysical tides.
\end{abstract}


\section{Introduction}

Tidal interactions determine the orbital and spin evolution of
astrophysical bodies when they orbit sufficiently close to one another.
Applications include close binary stars, extrasolar planetary systems
and the satellites of solar-system planets.

In many cases of interest the tidally forced bodies are fully or
partly convective.  In order to determine the rate of tidal evolution
it is therefore necessary to study the interaction between tides and
convection.  The response of a fluid body to tidal forcing generally
consists of two components: one is non-wavelike and of large scale,
while the other involves internal waves of smaller scale.  At the
simplest level the convection could be thought of as providing an
effective viscosity that damps the non-wavelike tidal disturbance and
provides a phase shift in the response.  As in the theory of
\citet{1880RSPT..171..713D}, the tidal torque and the rate of tidal
evolution are then directly proportional to this effective viscosity.
Convection may also play an important role in dissipating inertial
waves, which constitute the low-frequency wavelike response of
rotating convective zones of stars and giant planets
\citep[e.g.][]{2004ApJ...610..477O}.

As pointed out by \citet{1966AnAp...29..489Z} and
\citet{1977Icar...30..301G}, the effective viscosity estimated from
mixing-length theory ought to be reduced when the period of the tidal
disturbance is short compared to the characteristic timescale of the
convective motion.  The suppression factor has been the subject of
much debate, informed by observational constraints such as the
apparent rate of circularization of the orbits of close binary stars
\citep[][and references therein]{2005ApJ...620..970M}.
\citet{1966AnAp...29..489Z} suggested a less severe suppression, such
that the effective viscosity is proportional to the oscillation period
for short periods, which is in better agreement with these
observations.  \citet{1977Icar...30..301G} and
\citet{1977ApJ...211..934G} considered the multiscale nature of the
turbulent flow, and argued that the dominant contribution to the
effective viscosity at short periods comes from eddies with a
convective timescale comparable to the oscillation period.  For a
Kolmogorov spectrum, this argument gives a more powerful suppression,
such that the effective viscosity is proportional to the square of the
oscillation period for short periods.

While these authors relied on simple physical arguments and
order-of-magnitude estimates, \citet{1997ApJ...486..403G}, who also
provide a clear review of the controversy, introduced a more formal
procedure for determining the effective viscosity of a convective
flow, by considering the effect of a homogeneous oscillatory strain on
a turbulent velocity field.  Analytical progress was limited by the
complexity of the equations.  While their method relies on an
expansion in powers of the ratio of the oscillation period to the
convective timescale, it leads to a result in which the dominant
contribution comes from eddies for which these timescales are similar.
Their argument does, however, appear to rule out the hypothesis of
\citet{1966AnAp...29..489Z}.

More recently, numerical simulations of convection have been brought
to bear on this question.  \citet{2007ApJ...655.1166P} and
\citet{2009ApJ...704..930P} applied the procedure of
\citet{1997ApJ...486..403G} to the velocity fields obtained in
numerical simulations of convection in a deep layer, to deduce the
effective viscosity as a function of the tidal frequency.  In
\citet{2009ApJ...705..285P} an oscillatory forcing was introduced
directly into the convection simulation and the effective viscosity
was estimated by measuring the work done by this force.  The results
of these studies suggest that something closer to the prescription of
\citet{1966AnAp...29..489Z} may be appropriate, not for the reasons
originally suggested, but possibly because the power spectrum of the
convection is less steep than the Kolmogorov spectrum assumed by
\citet{1977Icar...30..301G}.

While the results of Penev and collaborators are of considerable
interest, the uncertainties in these calculations have not been
quantified.  Owing to the importance of this problem, we are motivated
to examine it from an independent viewpoint.  We have devised a
theoretical framework that allows us to study statistically
homogeneous flows; while this local viewpoint omits important global
aspects of stellar or planetary convection, it allows us to resolve
convection at high Rayleigh numbers and to quantify the uncertainties
due to turbulent noise.  We also present complementary analytical
results which shed new light on previous theoretical discussions.

The plan of this paper is as follows.  In Section~\ref{s:osb} we
introduce the homogeneous sheared coordinate system used in our
analytical and numerical calculations, and discuss the equations of
fluid dynamics in this system.  In Section~\ref{s:hf} we derive the
response of a fluid flow to oscillatory shear in the limits of low
amplitude and high frequency, and calculate the effective elasticity
and viscosity in various cases.  Direct numerical simulations of
convection in an oscillatory shearing box are reported and interpreted
in Section~\ref{s:dns}, followed by a summary and discussion of the
results.

\section{The oscillatory shearing box}

\label{s:osb}

\subsection{Motivation}

{At the present time it is not practical to compute a global
  model of an astrophysical body with turbulent convection and to
  measure its tidal response through the direct application of a
  time-dependent gravitational potential.  As noted in the
  introduction, the response of a fluid body to tidal forcing
  generally consists of a large-scale non-wavelike motion together
  with some internal waves.  The non-wavelike tide can be computed
  without difficulty in a linear approximation and consists of a
  time-dependent spheroidal deformation of the body.  From the local
  perspective of the smaller-scale convective motion, this tidal flow
  appears as a spatially homogeneous, approximately incompressible
  motion that is oscillatory in time.  We are interested, therefore,
  in determining the response of the convective motion to such a
  deformation and the additional stresses that result from this
  interaction.}

{A general, three-dimensional, spatially homogeneous,
  incompressible deformation can be represented by a traceless
  velocity gradient tensor $\bnabla\bu$ that is independent of
  position.  In this paper we consider a time-dependent simple shear
  of the form $u_y\propto x$.  Using a linear combination of simple
  shears with different orientations it is possible to compose an
  arbitrary velocity gradient tensor that is traceless (see
  Appendix~\ref{s:proof}).  Provided that we are working a linear
  regime, therefore, it is sufficient to measure the response to a
  time-dependent simple shear.}

\subsection{Sheared coordinates}

We initially consider an incompressible fluid in two dimensions,
satisfying the Navier--Stokes equations
\begin{equation}
  (\p_t+u_x\p_x+u_y\p_y)u_x=-\p_xp+\nu(\p_x^2+\p_y^2)u_x+\tilde f_x,
\end{equation}
\begin{equation}
  (\p_t+u_x\p_x+u_y\p_y)u_y=-\p_yp+\nu(\p_x^2+\p_y^2)u_y+\tilde f_y,
\end{equation}
\begin{equation}
  \p_xu_x+\p_yu_y=0,
\end{equation}
where $\bu$ is the velocity, $p$ is the pressure divided by the
density, $\nu$ is the kinematic viscosity and $\tilde\bff$ is a body force
per unit mass.

We suppose that the system is subject to a homogeneous time-dependent
shear.  We define the sheared coordinates
\begin{equation}
  x'=x,\qquad
  y'=y-a(t)x,\qquad
  t'=t.
\end{equation}
The dimensionless quantity $a$ is the shear, and $\dot
a=\rmd a/\rmd t$ is the shear rate.  The Jacobian determinant of the transformation is unity.  Partial derivatives transform
according to
\begin{equation}
  \p_x=\p_x'-a\p_y',\qquad
  \p_y=\p_y',\qquad
  \p_t=\p_t'-\dot ax\p_y'.
\end{equation}
We write
\begin{equation}
  u_x=v_x,\qquad
  u_y=v_y+\dot ax,
\end{equation}
so that $\bv$ represents the velocity relative to the shearing motion.
We continue to refer vector components to the original Cartesian
basis.

With these substitutions, the Navier--Stokes equations become
\begin{equation}
  [\p_t'+v_x(\p_x'-a\p_y')+v_y\p_y']v_x=-(\p_x'-a\p_y')p+\nu\Delta v_x+\tilde f_x,
\end{equation}
\begin{equation}
\begin{split}
  &[\p_t'+v_x(\p_x'-a\p_y')+v_y\p_y']v_y+\ddot ax'+\dot av_x=-\p_y'p\nonumber\\
  &\qquad+\nu\Delta v_y+\tilde f_y,
\end{split}
\end{equation}
\begin{equation}
  (\p_x'-a\p_y')v_x+\p_y'v_y=0,
\end{equation}
involving the Laplacian operator in unsheared coordinates,
\begin{equation}
\begin{split}
  \Delta&=\p_x^2+\p_y^2\nonumber\\
  &=(\p_x'-a\p_y')^2+\p_y'^2\nonumber\\
  &=\Delta'-2a\p_x'\p_y'+a^2\p_y'^2.
\end{split}
\end{equation}

These equations are spatially homogeneous (not involving $x'$ or $y'$
explicitly) except for the term $\ddot ax'$, which arises because a
spatially inhomogeneous force is required to maintain an accelerating
shear.  By writing
\begin{equation}
  \tilde f_x=f_x,\qquad
  \tilde f_y=\ddot ax'+f_y,
\end{equation}
i.e.\ by subtracting the inhomogeneous force, we restore spatial
homogeneity to the equations.  {In this representation, $\bff$
  is the body force that, if necessary, drives the (possibly
  turbulent) motion whose response we wish to measure, while $\ddot
  ax'\,\be_y$ is the force that drives the imposed large-scale shear.}

The property of spatial homogeneity results from the translational
invariance of the homogeneously sheared system, given that a Galilean
transformation easily removes the velocity shift associated with a
translation in the $x$ direction. It means that statistically
homogeneous turbulence, for example, is possible in this system.  It
also means that, for computational or analytical purposes, we can
impose periodic boundary conditions on $\bv$ and $p$ in sheared
coordinates.  When employing a spectral method, these variables can be
expanded in Fourier series in sheared coordinates, using basis
functions $\exp(\rmi k_x'x'+\rmi k_y'y')$.

An alternative approach, which leads to identical results, is to solve
the original equations in unsheared coordinates but to apply modified
periodic boundary conditions of the form
\begin{equation}
  p(L_x,y,t)=p(0,(y-a(t)L_x)\bmod L_y,t),
\end{equation}
\begin{equation}
  p(x,L_y,t)=p(x,0,t),
\end{equation}
and similarly for $\bv$ (not $\bu$), where $L_x$ and $L_y$ are the
dimensions of the shearing box.  The appropriate basis is then one
composed of shearing waves $\exp(\rmi k_x(t)x+\rmi k_yy)$, having
(quantized) time-dependent wavevectors with components
$k_x(t)=k_x'-a(t)k_y'$ and $k_y=k_y'$.  With these definitions,
$\exp(\rmi k_x(t)x+\rmi k_yy)=\exp(\rmi k_x'x'+\rmi k_y'y')$.

{Note that the oscillatory shear in our model is not imposed by
  the boundary conditions as such, but is driven by the inhomogeneous
  body force $\ddot ax'\,\be_y$ as described above.  Our model
  therefore provides a self-consistent local representation of any
  turbulent flow subject to an oscillatory deformation due to an
  external body force.}

Similar considerations apply to the Boussinesq equations in three
dimensions, or indeed to the equations of compressible convection.  In
the Boussinesq case the basic equations in unsheared coordinates are
\begin{equation}
  (\p_t+u_j\p_j)u_i=-\p_ip+\nu\Delta u_i+be_i+\tilde f_i,
\end{equation}
\begin{equation}
  (\p_t+u_i\p_i)b+N^2u_ie_i=\kappa\Delta b,
\end{equation}
\begin{equation}
  \p_iu_i=0,
\end{equation}
where $b$ is a buoyancy variable (proportional to the Eulerian entropy
perturbation multiplied by the gravitational acceleration), $\be$ is a
unit vector in the direction opposite to gravity, $N^2$ is the square
of the buoyancy frequency (negative in the convectively unstable case)
and $\kappa$ is the thermal diffusivity.  {(In the case of
  convection no additional body force $\bff$ is required to drive the
  motion.)}

The standard shearing box, as employed by, e.g.,
\citet{1981STIN...8131508R}, \citet{1988AJ.....95..925W} and
\citet{1995ApJ...440..742H}, corresponds to setting $a(t)\propto t$
and, if appropriate, adding rotation to the box.  In this case the
shear is inexorable and the shearing coordinates must be remapped
periodically for the purposes of numerical simulation.

In our case we consider $a(t)$ to oscillate sinusoidally and no
remapping is required.  As the frequency of the oscillation is varied,
we can choose either to keep the amplitude of $a$ the same, which
means that the maximum angle of the shear is fixed, or to scale the
amplitude of $a$ inversely with the frequency, which means that the
maximum angular velocity of the shear is fixed.  For the most part we
are interested in the regime in which the shear is very small (in
either sense), but if it is too small then its effects cannot be
measured reliably in a direct numerical simulation.

The quantity to be measured that is of greatest interest is the shear
stress associated with the fluid motions, i.e.\ the Reynolds stress
component
\begin{equation}
  -R_{xy}=-\langle v_xv_y\rangle,
\end{equation}
where the angle brackets denote a suitable averaging operation.  (The
physical Reynolds stress has an additional factor of the density of
the fluid.)  It is this stress that exchanges energy with the
large-scale shear.  The energy equation in sheared coordinates in the
absence of buoyancy forces is
\begin{equation}
  \p_t'(\half v_iv_i)+\p_iF_i=-\dot av_xv_y-\nu|\bnabla\times\bv|^2+f_iv_i,
\end{equation}
where $\bF$ is an appropriate energy flux density.  If the Reynolds
stress $-R_{xy}$ is positively correlated with the shear rate $\dot a$
then, like a viscous stress, it extracts energy from the large-scale
shear and imparts it to smaller-scale motions.  However, there is no
reason in principle why the correlation should not be negative, in
which case work is done on the large-scale shear at the expense of
either the body force or the decaying energy of the smaller-scale
flow.

\section{High-frequency response}

\label{s:hf}

\subsection{Linearized equations}

In this section we present analytical results on the response of fluid
motions to high-frequency shear of low amplitude.  We omit buoyancy
forces, which are probably inessential for this discussion, and
consider the Navier--Stokes equations in sheared coordinates,
\begin{equation}
\begin{split}
  &[\p_t'+v_j(\p_j'-a\delta_{j1}\p_y')]v_i+\dot av_x\delta_{i2}=-(\p_i'-a\delta_{i1}\p_y')p\nonumber\\
  &\qquad+\nu(\p_j'-a\delta_{j1}\p_y')(\p_j'-a\delta_{j1}\p_y')v_i+f_i,
\end{split}
\end{equation}
\begin{equation}
  (\p_i'-a\delta_{i1}\p_y')v_i=0,
\end{equation}
where $\delta_{ij}$ is the Kronecker delta and the subscripts $1$ and
$2$ refer to the $x$ and $y$ directions involved in the shear.  We now
carry out a linearization in the shear amplitude.  We consider a basic
flow that exists in the absence of shear, satisfying the equations
\begin{equation}
  (\p_t'+v_j\p_j')v_i=-\p_i'p+\nu\Delta'v_i+f_i,
\end{equation}
\begin{equation}
  \p_i'v_i=0.
\end{equation}
This flow might be freely decaying, or it might for example be a
steady (or statistically steady) flow sustained by a body force.
{(Since we have omitted buoyancy forces, we cannot consider
  convection as such in this section, but we can study laminar or
  turbulent flows driven by a body force.)}  If we assume that
$\p_i'f_i=0$, since only the solenoidal part of the force drives the
flow, then $p$ satisfies the Poisson equation
\begin{equation}
  \Delta'p=-(\p_i'v_j)\p_j'v_i.
\end{equation}
The Laplacian operator $\Delta'$ has a unique inverse $\Delta'^{-1}$
if, as we assume, the boundary conditions are periodic and the fields
have zero mean.  This inverse is also a self-adjoint operator.

The linearized equations are
\begin{equation}
\begin{split}
  &(\p_t'+v_j\p_j')\delta v_i+(\delta v_j\p_j'-av_x\p_y')v_i+\dot av_x\delta_{i2}\nonumber\\
  &\qquad=-\p_i'\delta p+a\delta_{i1}\p_y'p+\nu(\Delta'\delta v_i-2a\p_x'\p_y'v_i),
\end{split}
\end{equation}
\begin{equation}
  \p_i'\delta v_i-a\p_y'v_x=0,
\end{equation}
where $\delta v_i$ is the velocity perturbation induced at first order
by the shear, and $\delta p$ is the accompanying pressure
perturbation.  In general the linearized equations must be solved
numerically.

\subsection{Asymptotic analysis for high frequencies}

We now consider a limit in which the shear is oscillatory with high
frequency.  By this we mean that the timescale of the oscillations is
small compared to the circulation period of the streamlines of the
basic flow and the viscous timescale.  For multiscale flows such as
fully developed turbulence, the approximation considered here applies
when the oscillation period is short compared to the convective
timescale of the smallest eddies.  Alternatively, it can be viewed as
determining the response of those eddies for which the convective
timescale is long compared to the oscillation period.

We use the method of multiple scales and introduce a fast time
variable $T'=t'/\epsilon$, where $\epsilon\ll1$ is a small parameter
that characterizes the ratio of timescales.  The rapidity of the shear
is expressed by rewriting $a\mapsto a(T')$ and $\dot
a\mapsto\epsilon^{-1}\dot a$, where the new meaning of $\dot a$ is
$\rmd a/\rmd T'$.  The basic flow may vary with the ordinary time
variable $t'$.  We then expand
\begin{equation}
  \delta v_i=\delta v_{i0}+\epsilon\,\delta v_{i1}+\cdots,
\end{equation}
\begin{equation}
  \delta p=\epsilon^{-1}(\delta p_0+\epsilon\,\delta p_1+\cdots),
\end{equation}
in asymptotic series, where the quantities on the right-hand side
depend on $(\bx',t',T')$.  The $\delta p$ expansion is indexed in this
way for reasons that will become clear.  Since the basic flow does not
depend on $\epsilon$, which is a property of the perturbations only,
it is not expanded.

At leading order we find
\begin{equation}
  \p_T'\delta v_{i0}+\dot av_x\delta_{i2}=-\p_i'\delta p_0,
\label{dtdvi0}
\end{equation}
\begin{equation}
  \p_i'\delta v_{i0}-a\p_y'v_x=0.
\end{equation}
Roughly speaking, these equations describe an elastic response in
which (from the second component of equation~\ref{dtdvi0}) $\delta
v_{y0}\approx-av_x$, leading to a shear stress $-\langle v_x\delta
v_{y0}\rangle\approx a\langle v_x^2\rangle$ proportional to the shear
and in phase with it.  The reality is somewhat more complicated
because of the pressure terms and the constraint of incompressibility.
The pressure perturbation satisfies a Poisson equation,
\begin{equation}
  \Delta'\delta p_0=-2\dot a\p_y'v_x,
\end{equation}
obtained by eliminating $\delta\bv$ from the above two equations, and
so
\begin{equation}
  \p_T'\delta v_{i0}=2\dot a\p_i'\p_y'\Delta'^{-1}v_x-\dot av_x\delta_{i2}.
\end{equation}
The linearized shear stress at this order is
\begin{equation}
  -\delta R_{xy0}=-\langle v_x\delta v_{y0}+v_y\delta v_{x0}\rangle
\end{equation}
and satisfies
\begin{equation}
  \p_T'(-\delta R_{xy0})=\dot a\langle v_x^2-2(v_x\p_y'+v_y\p_x')\p_y'\Delta'^{-1}v_x\rangle.
\end{equation}
In terms of the tensor
\begin{equation}
  A_{ijkl}=\langle v_i\p_j'\p_k'\Delta'^{-1}v_l\rangle,
\end{equation}
we have
\begin{equation}
  \p_T'(-\delta R_{xy0})=\dot a(A_{1jj1}-2A_{1221}-2A_{2121}).
\label{dtrxy0}
\end{equation}
This tensor has the symmetry property $A_{ijkl}=A_{ikjl}$, which
follows immediately from its definition, and the contraction
properties $A_{ijkj}=A_{ijkk}=0$, which follow from $\p_i'v_i=0$.  The
further symmetry property $A_{ijkl}=A_{ljki}$ follows from integration
by parts, if the averaging operation includes a spatial average over a
periodic cell.  The quantity multiplying $\dot a$ on the right-hand
side of equation~(\ref{dtrxy0}) can be regarded as the effective
elastic modulus (shear modulus) of the flow (divided by the density of
the fluid).

At the next order we find
\begin{equation}
\begin{split}
  &\p_T'\delta v_{i1}+(\p_t'+v_j\p_j')\delta v_{i0}+(\delta v_{j0}\p_j'-av_x\p_y')v_i\nonumber\\
  &\qquad=-\p_i'\delta p_1+a\delta_{i1}\p_y'p+\nu(\Delta'\delta v_{i0}-2a\p_x'\p_y'v_i),
\end{split}
\end{equation}
\begin{equation}
  \p_i'\delta v_{i1}=0.
\end{equation}
We then obtain the Poisson equation
\begin{equation}
\begin{split}
  &\Delta'\delta p_1=-a(\p_t'+v_j\p_j')\p_y'v_x-2(\p_i'v_j)\p_j'\delta v_{i0}\nonumber\\
  &\qquad+a(\p_i'v_x)\p_y'v_i+a\p_x'\p_y'p+a\nu\Delta'\p_y'v_x,
\end{split}
\end{equation}
which can be inverted in principle.  The linearized shear stress at
this order is
\begin{equation}
  -\delta R_{xy1}=-\langle v_x\delta v_{y1}+v_y\delta v_{x1}\rangle
\end{equation}
and satisfies
\begin{equation}
\begin{split}
  &\p_T'(-\delta R_{xy1})=\langle v_x(\p_t'+v_j\p_j')\delta v_{y0}+v_y(\p_t'+v_j\p_j')\delta v_{x0}\nonumber\\
  &\qquad+v_x(\delta v_{j0}\p_j'-av_x\p_y')v_y+v_y(\delta v_{j0}\p_j'-av_x\p_y')v_x\nonumber\\
  &\qquad+(v_x\p_y'+v_y\p_x')\delta p_1-av_y\p_y'p\nonumber\\
  &\qquad-\nu v_x(\Delta'\delta v_{y0}-2a\p_x'\p_y'v_y)\nonumber\\
  &\qquad-\nu v_y(\Delta'\delta v_{x0}-2a\p_x'\p_y'v_x)\rangle.
\end{split}
\end{equation}
We substitute for $p$ and $\delta p_1$ from the Poisson equations that
they satisfy, using the fact that the inverse Laplacian is
self-adjoint, and integrating by parts in various places:
\begin{equation}
\begin{split}
  &\p_T'(-\delta R_{xy1})=\langle v_x(\p_t'+v_j\p_j')\delta v_{y0}+v_y(\p_t'+v_j\p_j')\delta v_{x0}\nonumber\\
  &\qquad+v_x(\delta v_{j0}\p_j'-av_x\p_y')v_y+v_y(\delta v_{j0}\p_j'-av_x\p_y')v_x\nonumber\\
  &\qquad-\Delta'^{-1}(\p_y'v_x-\p_x'v_y)[-a(\p_t'+v_j\p_j')\p_y'v_x\nonumber\\
  &\qquad-2(\p_i'v_j)\p_j'\delta v_{i0}+a(\p_i'v_x)\p_y'v_i+a\p_x'\p_y'p+a\nu\Delta'\p_y'v_x]\nonumber\\
  &\qquad-a(\p_i'v_j)(\p_j'v_i)\p_y'\Delta'^{-1}v_y-\nu v_x(\Delta'\delta v_{y0}-2a\p_x'\p_y'v_y)\nonumber\\
  &\qquad-\nu v_y(\Delta'\delta v_{x0}-2a\p_x'\p_y'v_x)\rangle.
\end{split}
\end{equation}
A further replacement of $p$ is required:
\begin{equation}
\begin{split}
  &\p_T'(-\delta R_{xy1})=\langle v_x(\p_t'+v_j\p_j')\delta v_{y0}+v_y(\p_t'+v_j\p_j')\delta v_{x0}\nonumber\\
  &\qquad+v_x(\delta v_{j0}\p_j'-av_x\p_y')v_y+v_y(\delta v_{j0}\p_j'-av_x\p_y')v_x\nonumber\\
  &\qquad-\Delta'^{-1}(\p_y'v_x+\p_x'v_y)[-a(\p_t'+v_j\p_j')\p_y'v_x\nonumber\\
  &\qquad-2(\p_i'v_j)\p_j'\delta v_{i0}+a(\p_i'v_x)\p_y'v_i+a\nu\Delta'\p_y'v_x]\nonumber\\
  &\qquad+a\Delta'^{-2}\p_x'\p_y'(\p_y'v_x+\p_x'v_y)(\p_i'v_j)(\p_j'v_i)\nonumber\\
  &\qquad-a(\p_i'v_j)(\p_j'v_i)\p_y'\Delta'^{-1}v_y-\nu v_x(\Delta'\delta v_{y0}-2a\p_x'\p_y'v_y)\nonumber\\
  &\qquad-\nu v_y(\Delta'\delta v_{x0}-2a\p_x'\p_y'v_x)\rangle.
\end{split}
\end{equation}
Since we have an expression for $\p_T'\delta v_{i0}$, we consider
\begin{equation}
\begin{split}
  &\p_T'^2(-\delta R_{xy1})=\langle v_x(\p_t'+v_j\p_j')\p_T'\delta v_{y0}+v_y(\p_t'+v_j\p_j')\p_T'\delta v_{x0}\nonumber\\
  &\qquad+v_x[(\p_T'\delta v_{j0})\p_j'-\dot av_x\p_y']v_y+v_y[(\p_T'\delta v_{j0})\p_j'-\dot av_x\p_y']v_x\nonumber\\
  &\qquad-\Delta'^{-1}(\p_y'v_x+\p_x'v_y)[-\dot a(\p_t'+v_j\p_j')\p_y'v_x\nonumber\\
  &\qquad-2(\p_i'v_j)\p_j'\p_T'\delta v_{i0}+\dot a(\p_i'v_x)\p_y'v_i+\dot a\nu\Delta'\p_y'v_x]\nonumber\\
  &\qquad+\dot a\Delta'^{-2}\p_x'\p_y'(\p_y'v_x+\p_x'v_y)(\p_i'v_j)(\p_j'v_i)\nonumber\\
  &\qquad-\dot a(\p_i'v_j)(\p_j'v_i)\p_y'\Delta'^{-1}v_y-\nu v_x(\Delta'\p_T'\delta v_{y0}-2\dot a\p_x'\p_y'v_y)\nonumber\\
  &\qquad-\nu v_y(\Delta'\p_T'\delta v_{x0}-2\dot a\p_x'\p_y'v_x)\rangle.
\end{split}
\end{equation}
Substituting for $\p_T'\delta v_{i0}$, we obtain
\begin{equation}
\begin{split}
  &\p_T'^2(-\delta R_{xy1})=\dot a\langle v_x(\p_t'+v_j\p_j')(2\p_y'\p_y'\Delta'^{-1}v_x-v_x)\nonumber\\
  &\qquad+v_y(\p_t'+v_j\p_j')(2\p_x'\p_y'\Delta'^{-1}v_x)\nonumber\\
  &\qquad+v_x[(2\p_j'\p_y'\Delta'^{-1}v_x)\p_j'-2v_x\p_y']v_y\nonumber\\
  &\qquad+v_y[(2\p_j'\p_y'\Delta'^{-1}v_x)\p_j'-2v_x\p_y']v_x\nonumber\\
  &\qquad-\Delta'^{-1}(\p_y'v_x+\p_x'v_y)[-(\p_t'+v_j\p_j')\p_y'v_x\nonumber\\
  &\qquad-2(\p_i'v_j)\p_j'(2\p_i'\p_y'\Delta'^{-1}v_x-v_x\delta_{i2})+(\p_i'v_x)\p_y'v_i\nonumber\\
  &\qquad+\nu\Delta'\p_y'v_x]+\Delta'^{-2}\p_x'\p_y'(\p_y'v_x+\p_x'v_y)(\p_i'v_j)(\p_j'v_i)\nonumber\\
  &\qquad-(\p_i'v_j)(\p_j'v_i)\p_y'\Delta'^{-1}v_y\nonumber\\
  &\qquad-\nu v_x(2\p_y'\p_y'v_x-\Delta'v_x-2\p_x'\p_y'v_y)\rangle.
\end{split}
\end{equation}
This can be rearranged as follows, after integration by parts:
\begin{equation}
\begin{split}
  &\p_T'^2(-\delta R_{xy1})=\dot a\langle(\p_t'v_x)(-v_x+\p_y'\p_y'\Delta'^{-1}v_x+\p_x'\p_y'\Delta'^{-1}v_y)\nonumber\\
  &\qquad+v_x(v_j\p_j')(2\p_y'\p_y'\Delta'^{-1}v_x-v_x)\nonumber\\
  &\qquad+v_y(v_j\p_j')(2\p_x'\p_y'\Delta'^{-1}v_x)\nonumber\\
  &\qquad+v_x[(2\p_j'\p_y'\Delta'^{-1}v_x)\p_j'-2v_x\p_y']v_y\nonumber\\
  &\qquad+v_y[(2\p_j'\p_y'\Delta'^{-1}v_x)\p_j'-2v_x\p_y']v_x\nonumber\\
  &\qquad-\Delta'^{-1}(\p_y'v_x+\p_x'v_y)[-(v_j\p_j')\p_y'v_x\nonumber\\
  &\qquad-2(\p_i'v_j)\p_j'(2\p_i'\p_y'\Delta'^{-1}v_x)+3(\p_y'v_j)\p_j'v_x]\nonumber\\
  &\qquad+\Delta'^{-2}\p_x'\p_y'(\p_y'v_x+\p_x'v_y)(\p_i'v_j)(\p_j'v_i)\nonumber\\
  &\qquad-(\p_i'v_j)(\p_j'v_i)\p_y'\Delta'^{-1}v_y\nonumber\\
  &\qquad-\nu v_x(\p_y'\p_y'v_x-\Delta'v_x-3\p_x'\p_y'v_y)\rangle.
\end{split}
\end{equation}
We define the tensors
\begin{equation}
  B_{ijkl}=\langle(\p_t'v_i)\p_j'\p_k'\Delta'^{-1}v_l\rangle,
\end{equation}
\begin{equation}
  C_{ijkl}=-\nu\langle v_i\p_j'\p_k'v_l\rangle,
\end{equation}
\begin{equation}
  D_{ijkl}=\langle v_iv_j\p_k'v_l\rangle,
\end{equation}
\begin{equation}
  D_{ijklmn}=\langle v_iv_j\p_k'\p_l'\p_m'\Delta'^{-1}v_n\rangle,
\end{equation}
\begin{equation}
  D_{ijklmnpq}=\langle v_iv_j\p_k'\p_l'\p_m'\p_n'\p_p'\Delta'^{-2}v_q\rangle,
\end{equation}
\begin{equation}
  E_{ijkl}=\langle v_m(\p_m'\p_n'\Delta'^{-1}\p_i'v_j)\p_n'\Delta'^{-1}\p_k'v_l\rangle,
\end{equation}
which satisfy $B_{ijkl}=B_{ikjl}$, $B_{ijkj}=0$,
$C_{ijkl}=C_{ikjl}=C_{ljki}$, $C_{ijkj}=0$ and various other
identities for $D$ and $E$.  We then find (after integration by parts)
\begin{equation}
\begin{split}
  &\p_T'^2(-\delta R_{xy1})=-\dot a(B_{1jj1}-B_{1221}-B_{1122}\nonumber\\
  &\qquad-C_{1221}+C_{1jj1}+3C_{1122}\nonumber\\
  &\qquad-2D_{1jj221}-2D_{2jj121}-3D_{1jj221}-3D_{1jj212}\nonumber\\
  &\qquad-D_{ijij1221}-D_{ijij1212}+D_{ijij22}\nonumber\\
  &\qquad+4E_{2121}+4E_{2112}).
\end{split}
\end{equation}

\subsection{Interpretation}

The two results obtained so far can be written in the forms
\begin{equation}
  \p_T'(-\delta R_{xy0})=\dot a\mathcal{G}_0,
\end{equation}
\begin{equation}
  \p_T'^2(-\delta R_{xy1})=-\dot a\mathcal{G}_1,
\end{equation}
where $\mathcal{G}_0$ and $\mathcal{G}_1$ are two coefficients, each
of which could be positive, negative or zero.  Given that this a
linear analysis, we may consider a complex shear
$a\propto\exp(-\rmi\omega t)$ with angular frequency
$\omega=O(\epsilon^{-1})$ and deduce that the shear stress in the
limit of high frequency is
\begin{equation}
  -\delta R_{xy}=a\left[\mathcal{G}_0-\f{\rmi\mathcal{G}_1}{\omega}+O(\epsilon^2)\right].
\label{g0g1}
\end{equation}
The term $\mathcal{G}_0$ represents an ideal elastic response, while
the term $\mathcal{G}_1$ represents an imperfection of the elasticity
associated with dissipation.  For comparison, the shear stress of a
viscous fluid is $\nu\dot a=-\rmi\omega\nu a$.  Therefore the
effective kinematic viscosity of the flow at high frequencies is
$\mathcal{G}_1/\omega^2$.

The calculation of \citet{1997ApJ...486..403G} corresponds only to the
leading order of the above expansion.  They do not mention the elastic
stress but focus on the dissipation rate at leading order in a
periodic strain, given by their equation~(25).  As they point out,
this expression evaluates to zero; this is consistent with our
analysis because there is no dissipation associated with a perfect
elastic stress.  They obtain a non-zero result at this order by
manipulating the apparent singularity in their expression at zero
frequency.  While this procedure may produce a meaningful result, it
is not really justified because the preceding steps have assumed a
separation of timescales between the tide and the convection; the
zero-frequency pole signals the breakdown of their approximation
scheme and its resolution is not straightforward.  In contrast, our
calculation of $\mathcal{G}_1$ and the effective viscosity in the
high-frequency limit is based on a systematic asymptotic expansion.

\subsection{Evaluation for statistically isotropic flows}

Although convective flows are naturally anisotropic, analytical
progress is easiest when the flow is assumed to have isotropic
properties when spatially averaged.  In this case $A_{ijkl}$ must be
an isotropic tensor, i.e.\
\begin{equation}
  A_{ijkl}=A_1\delta_{ij}\delta_{kl}+A_2\delta_{ik}\delta_{jl}+A_3\delta_{il}\delta_{jk}.
\end{equation}
The symmetry and contraction properties then imply $A_2=A_1$ and
$A_3=-(d+1)A_1$, where $d$ is the number of spatial dimensions (of
course $d=3$, but the case $d=2$ is also of at least theoretical
interest), and so
\begin{equation}
  A_{ijkl}=\f{2K}{d(d-1)(d+2)}[(d+1)\delta_{il}\delta_{jk}-\delta_{ij}\delta_{kl}-\delta_{ik}\delta_{jl}],
\end{equation}
where the mean kinetic energy density $K$ is given by
\begin{equation}
  2K=\langle v_iv_i\rangle=A_{ijji}=-d(d-1)(d+2)A_1.
\end{equation}
In this isotropic case we then obtain
\begin{equation}
  \mathcal{G}_0=\f{2(d-2)(d+1)}{d(d-1)(d+2)}K.
  \label{eq:iso_elastic}
\end{equation}
The effective elastic modulus is positive for a three-dimensional flow
but vanishes in two dimensions under the assumption of isotropy.

In the isotropic case it can be shown (see Appendix~\ref{s:triple})
that the triple-correlation tensor $D$ vanishes identically.  $E$ also
vanishes when the identity $E_{ijkl}=-E_{klij}$ (which follows from
integration by parts) is combined with the general form of a
fourth-rank isotropic tensor.  The tensors $B$ and $C$ have the form
\begin{equation}
  B_{ijkl}=B[(d+1)\delta_{il}\delta_{jk}-\delta_{ij}\delta_{kl}-\delta_{ik}\delta_{jl}],
\end{equation}
\begin{equation}
  C_{ijkl}=C[(d+1)\delta_{il}\delta_{jk}-\delta_{ij}\delta_{kl}-\delta_{ik}\delta_{jl}].
\end{equation}
Furthermore, the energy equation of the basic flow,
\begin{equation}
  \langle(\p_t'v_i)v_i\rangle=\nu\langle v_i\Delta'v_i\rangle+\langle f_iv_i\rangle,
\end{equation}
implies
\begin{equation}
  B_{ijji}+C_{ijji}=\langle f_iv_i\rangle=P,
\end{equation}
the power input per unit volume (or area), and so
\begin{equation}
  d(d-1)(d+2)(B+C)=P
\end{equation}
for isotropic statistics.  In this case
\begin{equation}
  \mathcal{G}_1=B(d^2-2)+C(d^2-6).
  \label{eq:iso_viscous}
\end{equation}
This corresponds to a viscous response, with the effective viscosity
being inversely proportional to frequency-squared, for high-frequency
oscillatory shear.  The effective viscosity coefficient is
$\mathcal{G}_1/\omega^2$.  In the case $B=0$, when the flow is
maintained steadily against viscous dissipation by the body force, the
effective viscosity is positive in three dimensions and negative in
two dimensions.  In the case $P=0$, when the flow is freely decaying
($-B=C>0$), the effective viscosity is negative in either three or two
dimensions.

\subsection{Evaluation for ABC flows}

A widely studied class of incompressible fluid flows in a periodic
domain is provided by the ABC flow \citep[named after Arnol'd,
Beltrami and Childress;][]{1965Arnol'd,1987JFM...180..557G}
\begin{equation}
  \bv=\left(\begin{matrix}A\sin kz'+C\cos ky'\\B\sin kx'+A\cos kz'\\C\sin ky'+B\cos kx'\end{matrix}\right)
\end{equation}
in a cube of length $2\pi/k$.  This velocity field has the Beltrami
property $\bnabla'\times\bv=k\bv$, so nonlinearity is absent;
$\bv\cdot\bnabla'\bv$ is balanced by a pressure gradient.  If the flow
is unforced, $A$, $B$ and $C$ decay proportionally to $\exp(-\nu k^2
t)$.  Alternatively the flow can be maintained against dissipation by
supplying a body force $\bff=\nu k^2\bv$.  The most widely studied
example has $A=B=C$.

The response coefficients are easily evaluated as
\begin{equation}
  \mathcal{G}_0\half(A^2-C^2),
\end{equation}
\begin{equation}
  \mathcal{G}_1=\half A(\dot A+\nu k^2A).
\end{equation}
Therefore the elasticity can be positive, negative or zero depending
on the anisotropy of the flow and its orientation relative to the
shear.  The effective viscosity $\mathcal{G}_1/\omega^2$ at leading
order vanishes for a freely decaying flow but is positive (and
inversely proportional to frequency-squared) for a forced flow,
assuming that $A\ne0$.

In fact, probably all analytical examples of three-dimensional fluid
flows lack genuine nonlinearity, in the sense that, if they are
expanded in a Fourier basis with wavenumbers $\bk$, there are no
non-empty triads of interacting components satisfying
$\bk_1+\bk_2+\bk_3=\mathbf{0}$.  Such triads tend to produce a cascade
of energy to larger wavenumbers in the manner of hydrodynamic
turbulence.  If there are no non-empty triads then the
triple-correlation tensors $D$ and $E$ again vanish, and the only
contributions to $\mathcal{G}_1$ come from the time-dependence of the
flow (the tensor $B$) and the viscous terms (the tensor $C$).  Both of
these effects may be small if the viscosity is small.

The ABC flow is stable only for sufficiently small Reynolds number.
More realistically, in a typical flow at high Reynolds number there
will be a turbulent cascade involving strong triad interactions,
meaning that the tensors $D$ and $E$ may be significant (although
apparently not in isotropic turbulence).  The tensors $B$ and $C$ will
also be enhanced by the turbulent cascade.  Numerical simulations are
required to access this regime.

{We note that the `eddy viscosity' of the $A=B=C$ flow has been
  calculated, as a function of Reynolds number, by
  \citet{1995JFM...288..249W}; see also references therein for related
  studies.  However, the nature of their calculation is different;
  using multiscale techniques, they determine the behaviour of
  \emph{very slow} large-scale deformations of the cellular flow, and
  deduce that a large-scale instability occurs through the appearance
  of a negative eddy viscosity.  In contrast, our analysis determines
  the response of a flow to an imposed large-scale deformation of high
  frequency.}

\section{Direct numerical simulations of convection in an oscillatory
  shearing box}

\label{s:dns}

\subsection{Numerical setup}

We have implemented the oscillatory shearing box in the SNOOPY code, a
3D spectral code solving the equations of incompressible or Boussinesq
(magneto)hydrodynamics using a Fourier representation
\citep{2005A&A...444...25L,2007MNRAS.378.1471L,2010MNRAS.404L..64L}.
We assume that an external force creates an oscillatory shear with
$\dot a=-S\cos(\Omega t)$, where $S$ is the maximum shear rate and
$\Omega$ is the tidal frequency. As in previous sections, {the
  shearing motion is in the $y$ direction, with a linear dependence on
  $x$, i.e.\ $-S\cos(\Omega t)x\,\be_y$, and the force that drives it
  is $\Omega S\sin(\Omega t)x\,\bm{e}_y$.}

In addition to this {imposed shearing motion}, convection is
driven by applying uniform gravity and an unstable entropy gradient,
within the Boussinesq approximation. {(No additional body force
  is required to drive the motion.)} By setting $N^2=-1$, we adopt a
unit of time related to the unstable stratification. We test two
different configurations: convection in the shearwise direction
($\bm{e}=\bm{e}_x$) and convection in the spanwise direction
($\bm{e}=\bm{e}_z$, where $\bm{e}$ is the direction of stratification
defined in Section~\ref{s:osb}).

The aspect ratio of the periodic box is adjusted according to the
direction of stratification, with $L_x\times L_y\times L_z=2\times
1\times 1$ when $\bm{e}=\bm{e}_x$ and $1\times 1\times 2$ when
$\bm{e}=\bm{e}_z$.  Having a box elongated in the direction of
stratification allows `elevator modes' to break up more easily through
secondary instabilities; otherwise, these tend to dominate the
convection at moderate Rayleigh numbers \citep{2010MNRAS.404L..64L}.
The resolution is $128$ collocation points per unit of length.

To control dissipation on small scales, we introduce an explicit
kinematic viscosity $\nu$ and thermal diffusivity $\kappa$, with
Prandtl number $\nu/\kappa=1$ for simplicity. The value of the
diffusion coefficients is set according to the Rayleigh number
$Ra=|N^2|L_y^4/\nu\kappa$. In our setup, convection starts when
$Ra>(2\pi)^4\approx 1559$. In the following we will consider
simulations exhibiting fully turbulent convection, with $Ra=4\times
10^6$.

\begin{figure}
  \centerline{\includegraphics[width=0.78\linewidth]{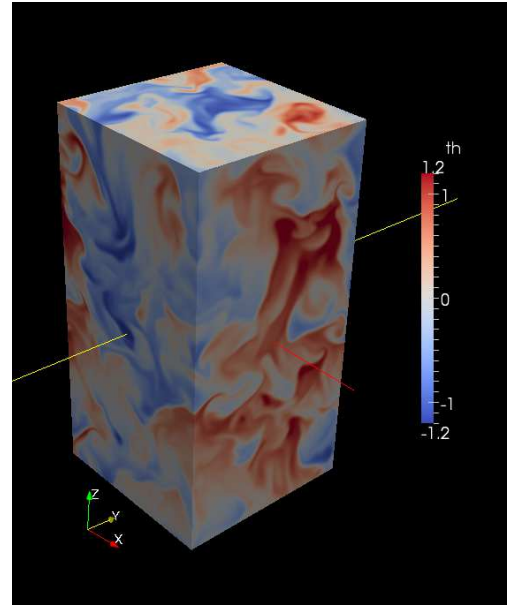}}
  \centerline{\includegraphics[width=1.0\linewidth]{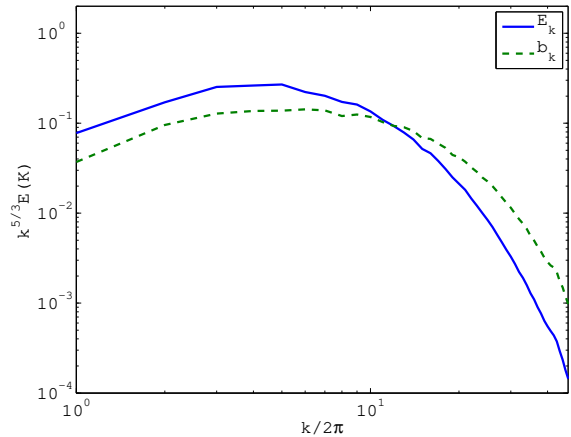}}
  \caption{Simulation snapshot of thermal fluctuations (top) and
    energy spectra (bottom) at $Ra=4\times 10^6$ in the case without
    any shear.  We show the kinetic energy spectrum ($E_k$) and the
    temperature fluctuation spectrum ($b_k$).}
  \label{Fig:noshear}
\end{figure}

In order to avoid any artefact of the initial conditions, we initiate
our simulations with noise at the largest scales and we let turbulence
evolve \emph{without any shear} for 100 turnover times (i.e.\ from
$t=-100 |N|^{-1}$ to $t=0$). The spectrum of the turbulence we obtain
and a typical snapshot are shown in Fig.~\ref{Fig:noshear}.  Once a
quasi-stationary turbulent state is reached, we switch on a weak
oscillatory shear and start measuring the Reynolds stress
$-R_{xy}(t)=-\langle v_x v_y\rangle$, where $\langle \cdot \rangle$
denotes a volume average over the box. Such a simulation has to be
continued for an integration time $\Delta T$ of hundreds of turnover
times in order to reduce the impact of turbulent noise on the
measurements.

\subsection{Turbulent viscosity: definitions and simple models}

Traditionally, the turbulent viscosity is associated with a simple
closure formula relating the Reynolds stress to the rate of strain; in
our system,
\begin{equation}
  -R_{xy}=\nu_\rmt\dot a.
\end{equation}
This expression is based on the assumptions that the relationship
between stress and rate of strain is linear and instantaneous.  In the
case of rapid oscillatory shear, the assumption of linearity is
reasonable provided that the shear is of sufficiently small amplitude.
However, the assumption of instantaneity is generally not valid and we
should allow the Reynolds stress to be linearly related to $\dot a(t)$
through an integral over its previous history.  In the Fourier domain,
this relationship (a convolution) reduces to a multiplication:
\begin{equation}
  -\widetilde{R}_{xy}(\omega)=\nu_\rmt(\omega)\widetilde{\dot a}(\omega),
\label{nut1}
\end{equation}
where $\widetilde{\cdot}$ denotes the Fourier transform in time. In
this expression $\nu_\rmt(\omega)$ is a complex function of frequency,
rather than a real constant, indicating that the Reynolds stress and
the rate of strain are generally out of phase, and that the
relationship is frequency-dependent.  Since the relationship involves
real-valued functions in the temporal domain, $\nu_\rmt$ has the
Hermitian symmetry $\nu_\rmt(-\omega)=[\nu_\rmt(\omega)]^*$ for real
$\omega$.  With our choice of $\dot a=-S\cos(\Omega t)$, we have
\begin{equation}
  \widetilde{R}_{xy}(\omega)=\nu_\rmt(\omega)S\,\pi[\delta(\omega-\Omega)+\delta(\omega+\Omega)].
\label{nut2}
\end{equation}
This makes it possible to measure $\nu_\rmt(\Omega)$ from the time
series $R_{xy}(t)$.  In reality, the delta function is replaced by a
peak of finite height and non-zero width because of the finite
integration time; alternatively, it becomes a Kronecker delta in a
discrete Fourier transform.  Furthermore, noise is present because
there are turbulent fluctuations in $R_{xy}$ even in the absence of
shear.  The integration time must be sufficiently long to allow the
signal to be detected above the noise.

A simple closure model for the Reynolds stress that can be compared
with the results of numerical simulations is a viscoelastic model,
which reflects the fact that the turbulent stress cannot
instantaneously produce a viscous response to a time-dependent shear
rate.  [Viscoelastic models for magnetohydrodynamic turbulence in
astrophysical discs have been discussed by
\citet{2001MNRAS.325..231O}.]  We suppose that the convective
turbulence contains a number of viscoelastic components, having
effective elastic moduli $c_j$ and relaxation times $1/\gamma_j$,
giving rise to the complex turbulent viscosity
\begin{equation}
  \nu_\rmt(\omega)=\sum_j\frac{c_j}{\gamma_j-\rmi\omega}.
\end{equation}
In this expression, $c_j$ and $\gamma_j$ are real, but we do not
require $c_j>0$, allowing for the possibility that negative
$\mathrm{Re}[\nu_\rmt(\omega)]$ may appear for some ranges of
frequency.  For example, the case of constant shear rate ($\omega=0$)
is known to produce negative viscosity for low values of $Ra$ in
protoplanetary disc convection, which involves rotation as well as
shear \citep{2010MNRAS.404L..64L}.  However, in order to have a
physical relaxation for all possible input signals, we require
$\gamma_j>0$, indicating that the system we describe is stable for any
time-dependent shear.  In other words, since $\nu_\rmt(\omega)$ in
equation~(\ref{nut1}) derives from a causal integral relationship, it
should be analytic in the upper half-plane.

This model shares several properties with the analytical results
derived in Section~\ref{s:hf}. In the high-frequency limit
$\omega\to\infty$, the model gives
\begin{equation}
  \nu_t=\rmi\omega^{-1}\sum_jc_j+\omega^{-2}\sum_jc_j\gamma_j+O(\omega^{-3}).
\end{equation}
This behaviour is equivalent to equation~(\ref{g0g1}) obtained for the
high-frequency response of an arbitrary flow, if
$\mathcal{G}_0=\sum_jc_j$ (the high-frequency elastic modulus) and
$\mathcal{G}_1=\sum_jc_j\gamma_j$.  (On the other hand, the viscosity
obtained in the low-frequency limit $\omega\to0$ is
$\sum_jc_j/\gamma_j$.)  As we will see later, the same type of
dependency is found in numerical simulations.

In this paper, we will assume that only two viscoelastic components
are present. This can be seen as a computational limitation, since
numerical simulations cannot probe very high frequencies which are
usually associated with scales below the grid size. In the following,
we will therefore consider the simplified model
\begin{equation}
\label{eq:num_closure}
  \nu_\rmt(\omega)=\frac{c_1}{\gamma_1-\rmi\omega}+\frac{c_2}{\gamma_2-\rmi\omega}.
\end{equation}

\subsection{Measuring the turbulent viscosity}

\begin{figure*}
  \centering
  \includegraphics[width=0.45\linewidth]{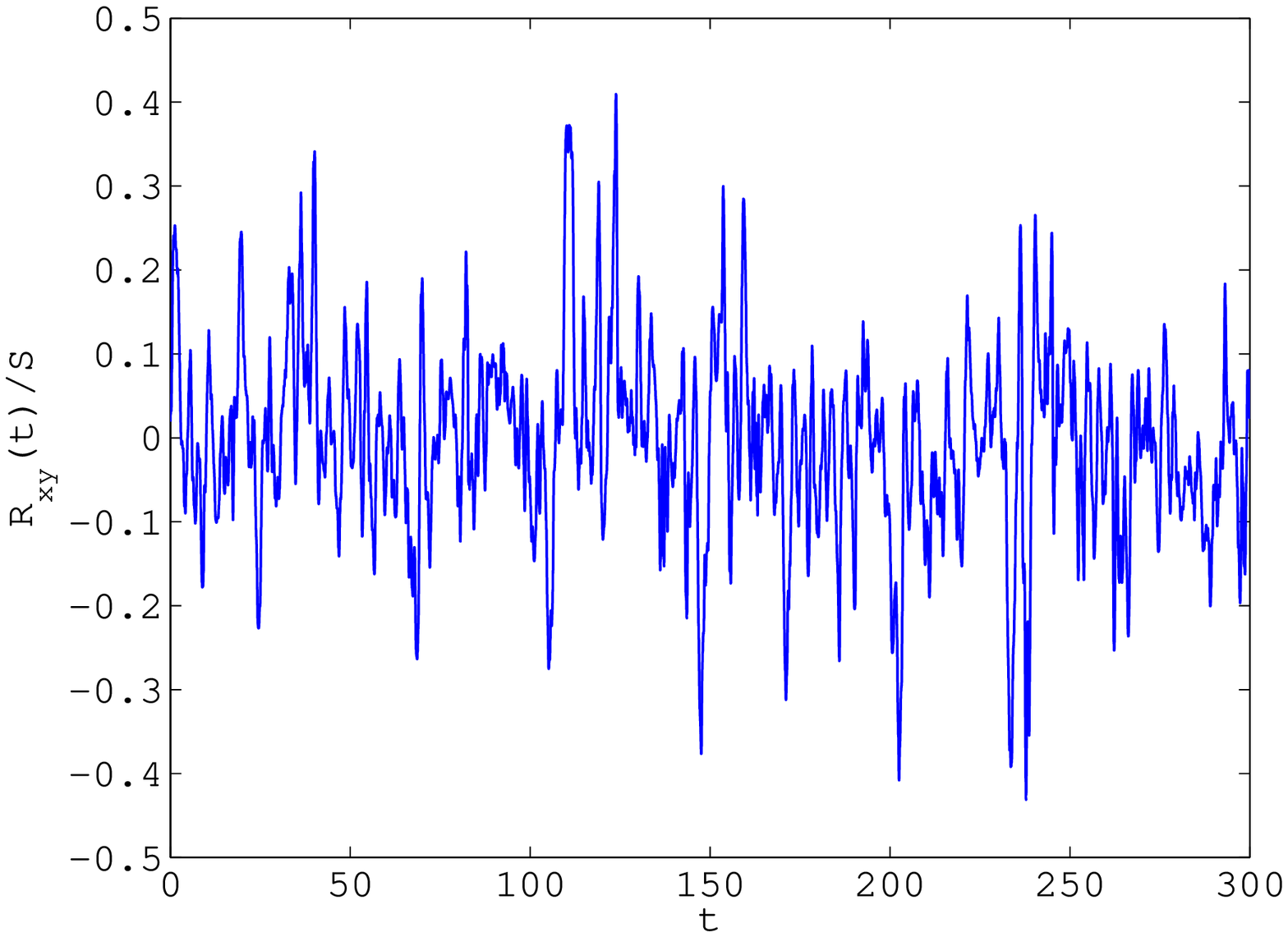}
  \includegraphics[width=0.45\linewidth]{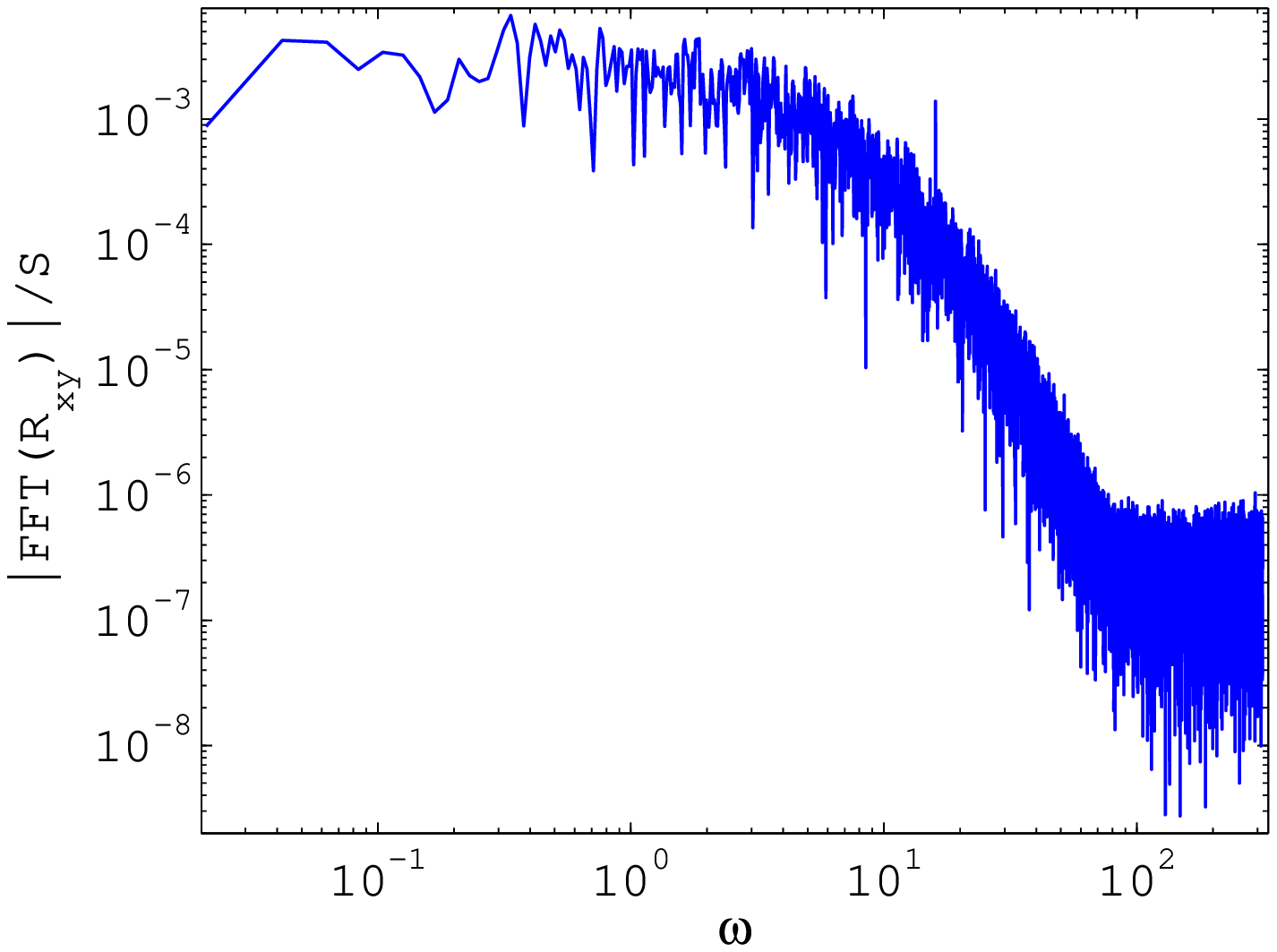}
  \caption{Time history of the Reynolds stress (left) and Fourier
    transform of the Reynolds stress (right) with an oscillatory shear
    of frequency $\Omega=16$ and amplitude $S=0.1$. The periodic
    signal cannot be seen in the time series. The Fourier transform
    exhibits a peak at the excitation frequency $\omega=\Omega$.}
  \label{Fig:RunExample}
\end{figure*}

A typical example of the time series of the Reynolds stress and its
Fourier transform are shown in Fig.~\ref{Fig:RunExample}. In order to
reduce the aliasing of low frequencies into the high-frequency domain
due to the finite integration time, we have applied a Hanning window
to the time series before computing the Fourier transform. As is
evident from the time series, the turbulent convection produces large
fluctuations in the Reynolds stress which dominate the response to the
oscillatory shear. Nevertheless, it is possible to extract useful
information by looking at the temporal spectrum of the stress.  In
particular, a spike is clearly visible at $\omega=\Omega$, which
indicates that the turbulent flow is producing a detectable response
to our forcing.  With the complete Fourier transform of the Reynolds
stress, it is possible to measure the turbulent viscosity at the
forcing frequency using equation~(\ref{nut2}). Moreover, evaluating
the noise level in the vicinity of the spike allows us to estimate the
uncertainty in the measurement of the turbulent viscosity.

It should be noted that the signal-to-noise ratio (defined as the
ratio of the amplitude of the spectral spike to the amplitude of the
surrounding noise; see Fig.~\ref{Fig:RunExample}, right panel) should
decay as $\Delta T^{-1/2}$.  Therefore, if the oscillatory signal (the
spectral spike) is too weak to be detected in the Fourier transform,
one can in principle increase the integration time to reduce the
signal-to-noise ratio, but this is a computationally expensive
procedure.

\subsection{Numerical results}

\begin{figure*}
  \centering
  \includegraphics[width=0.45\linewidth]{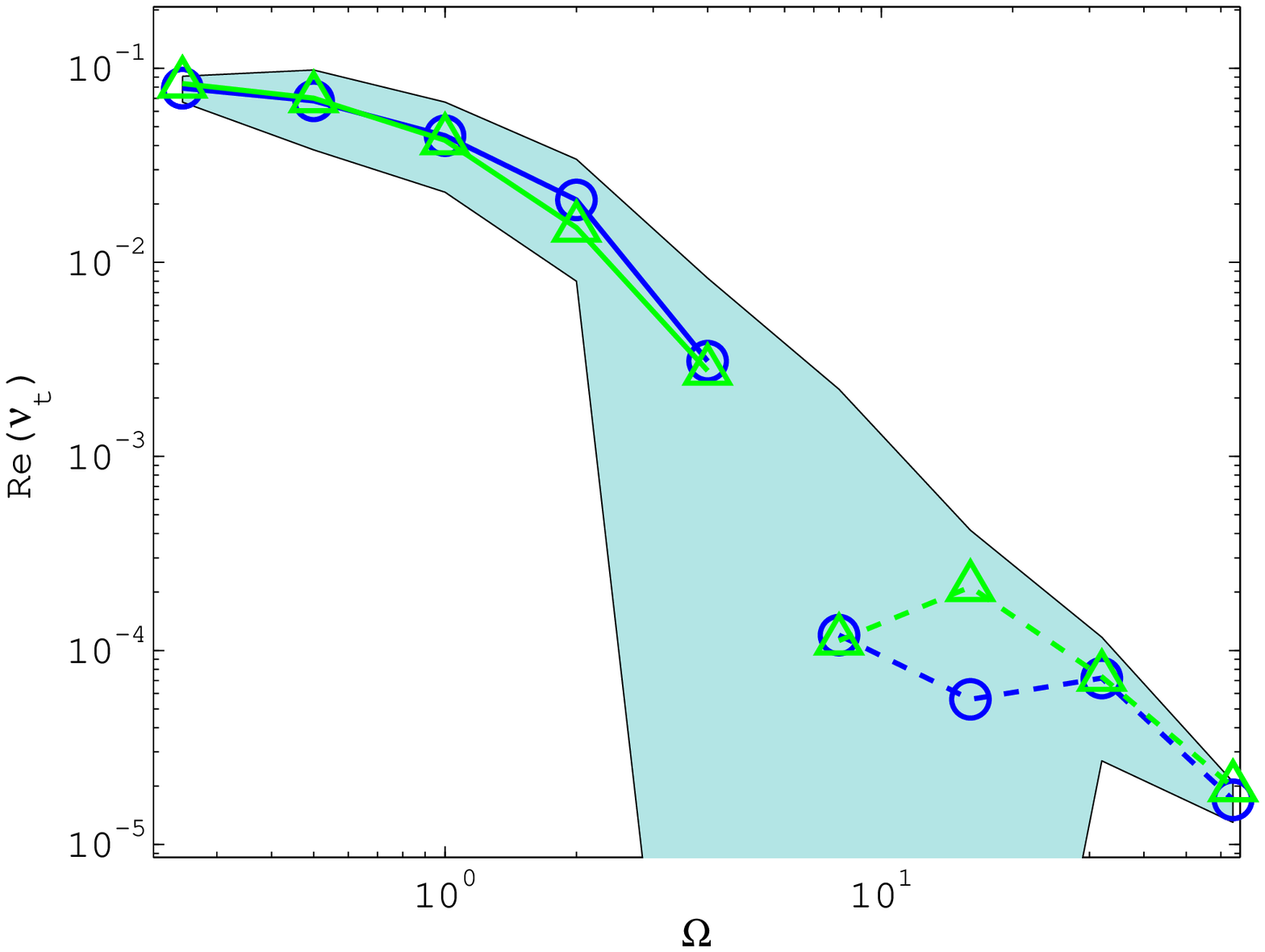}
  \includegraphics[width=0.45\linewidth]{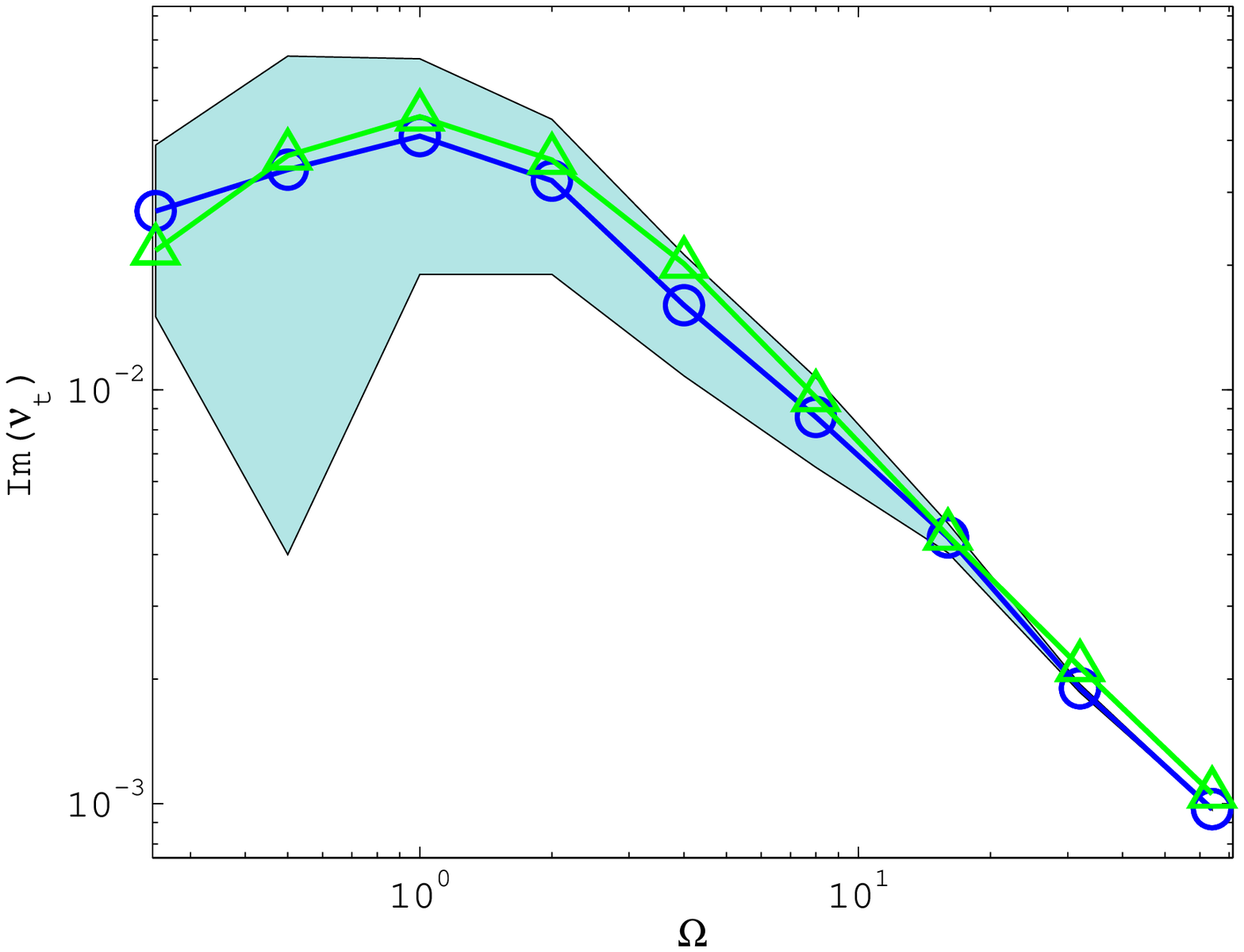}
  \caption{Turbulent viscosity of shearwise convection, versus the
    angular oscillation frequency in units of $|N|$.  The numerical
    measurements are shown in blue (open circles) and the best fit of the closure
    model (\ref{eq:num_closure}) is shown in green (triangles).  Negative values
    are connected by a dashed line.  Uncertainties due to turbulent
    noise are shown as a shaded region.}
  \label{Fig:shearwise}
\end{figure*}

\begin{figure*}
  \centering
  \includegraphics[width=0.45\linewidth]{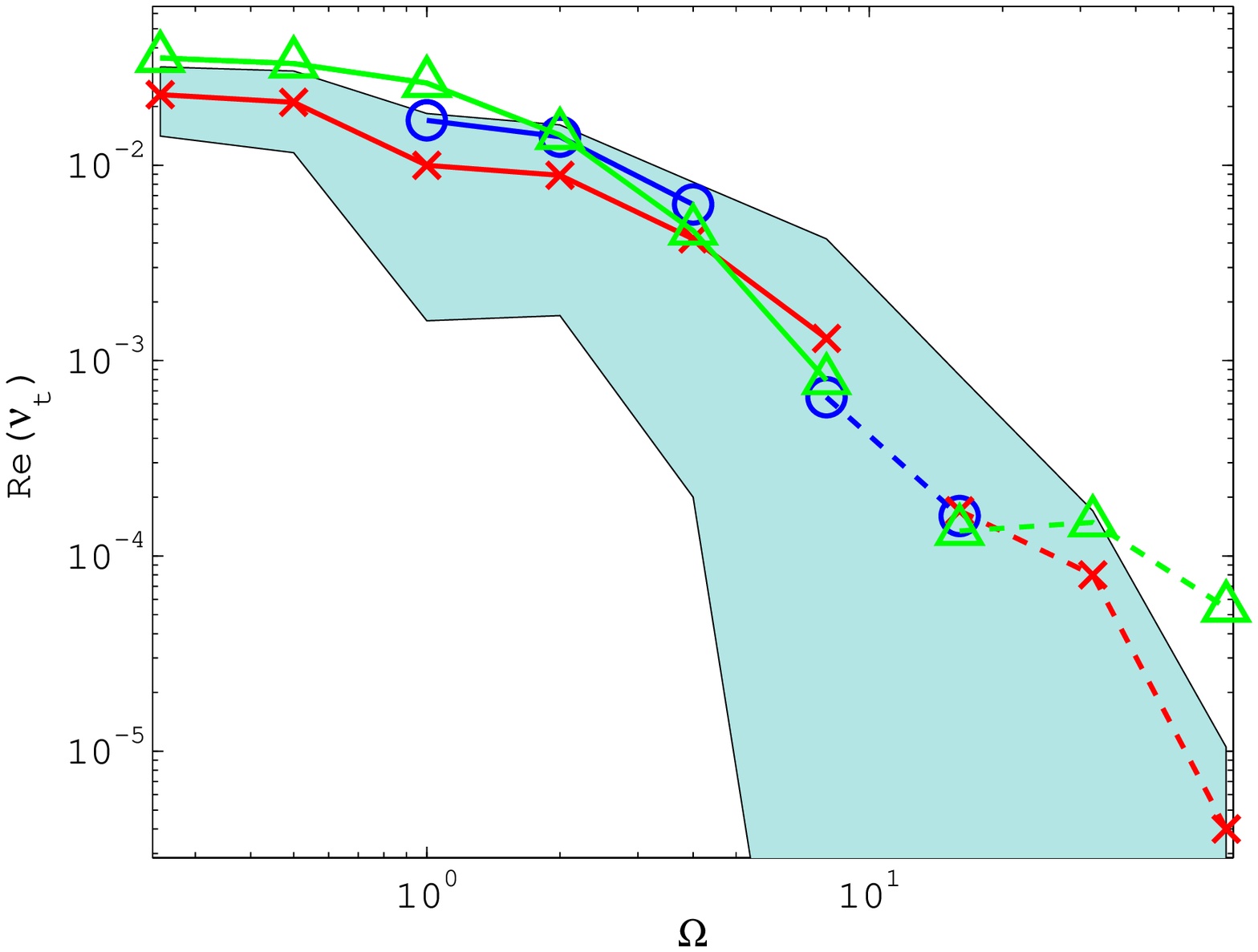}
  \includegraphics[width=0.45\linewidth]{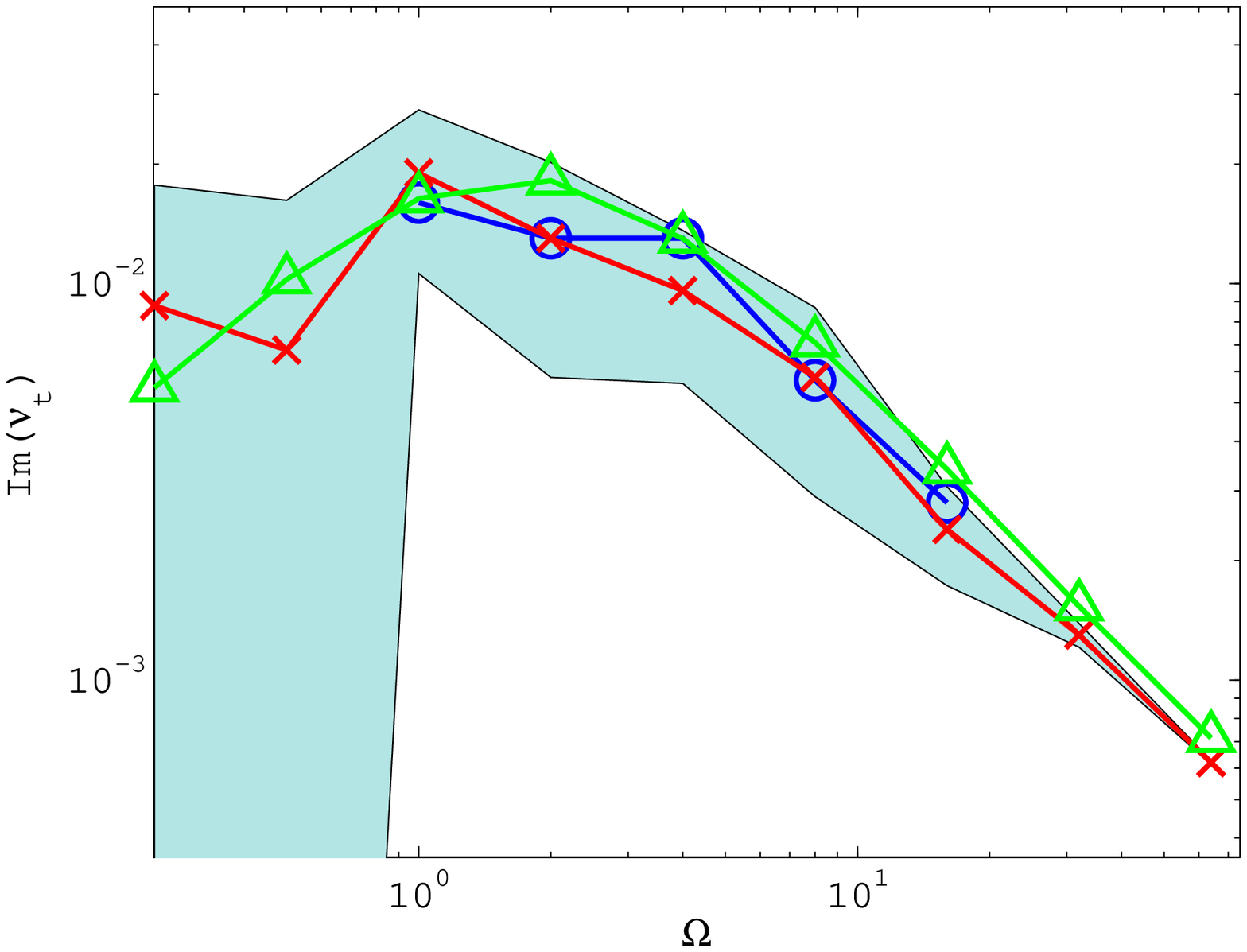}
  \caption{Turbulent viscosity of spanwise convection, versus the
    angular oscillation frequency in units of $|N|$. The numerical
    measurements are shown in blue (open circles; $S=0.1$) and red (crosses; $S=0.05$) and
    the best fit of the closure model (\ref{eq:num_closure}) is shown
    in green (triangles). Negative values are connected by a dashed line.
    Uncertainties due to turbulent noise are shown as a shaded region.}
  \label{Fig:spanwise}
\end{figure*}

The results of the simulations performed in this work are shown in
Table~\ref{tab:results}.  They are also plotted in
Fig.~\ref{Fig:shearwise} for shearwise convection and in
Fig.~\ref{Fig:spanwise} for spanwise convection. Despite the present
of significant noise in some of the results, several deductions can be
made from these simulations:

\begin{table*}
\begin{tabular}{ccccccrrrr}
\hline
$Ra$ & $\bm{e}$ direction & $\Delta T$ & $\Omega$ & $S$ & $K$  &    $\mathrm{Re}(\nu_\rmt)$      &     $\mathrm{Im}(\nu_\rmt)$    & error\\
\hline
$4\times 10^6$& $ x$       & 500           &      0.25     & 0.1   &  $2.1\times 10^{-1}$  & $7.9\times10^{-2}$   & $2.7\times 10^{-2}$ &   $1.2\times 10^{-2}$\\
$4\times 10^6$& $ x$       & 300           &      0.5       & 0.1   &  $2.2\times 10^{-1}$  & $6.8\times10^{-2}$   & $3.4\times 10^{-2}$ &   $3.0\times 10^{-2}$\\
$4\times 10^6$& $ x$       & 300           &      1.0       & 0.1   &  $2.1\times 10^{-1}$  & $4.5\times10^{-2}$   & $4.1\times 10^{-2}$ &   $2.2\times 10^{-2}$\\
$4\times 10^6$& $ x$       & 300           &      2.0       & 0.1   &  $2.2\times 10^{-1}$  & $2.1\times10^{-2}$   & $3.2\times 10^{-2}$ &   $1.3\times 10^{-2}$\\
$4\times 10^6$& $ x$       & 300           &      4.0       & 0.1   &  $2.1\times 10^{-1}$  & $3.1\times10^{-3}$   & $1.6\times 10^{-2}$ &   $5.2\times 10^{-3}$\\
$4\times 10^6$& $ x$       & 300           &      8.0       & 0.1   &  $2.2\times 10^{-1}$  & $-1.2\times10^{-4}$   & $8.6\times 10^{-3}$ &  $2.1\times 10^{-3}$\\
$4\times 10^6$& $ x$       & 300           &      16.0     & 0.1   &  $2.2\times 10^{-1}$  & $-5.6\times10^{-5}$   & $3.4\times 10^{-3}$ &  $3.6\times 10^{-4}$\\
$4\times 10^6$& $ x$       & 300           &      32.0     & 0.1   &  $2.2\times 10^{-1}$  & $-7.2\times10^{-5}$   & $1.9\times 10^{-3}$ &  $4.5\times 10^{-5}$\\
$4\times 10^6$& $ x$       & 300           &      64.0     & 0.1   &  $2.2\times 10^{-1}$  & $-1.7\times10^{-5}$   & $9.7\times 10^{-4}$ &  $4.0\times 10^{-6}$\\
\hline
$4\times 10^6$& $ z$       & 300           &      1.0      & 0.1   &  $2.3\times 10^{-1}$  &$1.7\times10^{-2}$   & $1.6\times 10^{-2}$ &   $5.6\times 10^{-3}$\\
$4\times 10^6$& $ z$       & 300           &      2.0      & 0.1   &  $2.1\times 10^{-1}$  &$1.4\times10^{-2}$   & $1.3\times 10^{-2}$ &   $4.6\times 10^{-3}$\\
$4\times 10^6$& $ z$       & 300           &      4.0      & 0.1   &  $2.2\times 10^{-1}$  &$6.3\times10^{-3}$   & $1.3\times 10^{-2}$ &   $3.4\times 10^{-3}$\\
$4\times 10^6$& $ z$       & 300           &      8.0      & 0.1   &  $2.2\times 10^{-1}$  &$-6.5\times10^{-4}$   & $5.7\times 10^{-3}$ &   $1.4\times 10^{-3}$\\
$4\times 10^6$& $ z$       & 300           &      16.0    & 0.1   &  $2.2\times 10^{-1}$  &$-1.6\times10^{-4}$   & $2.8\times 10^{-3}$ &   $4.0\times 10^{-4}$\\
\hline
$4\times 10^6$& $ z$       & 700           &      0.25    & 0.05  & $2.2\times 10^{-1}$  &$2.3\times10^{-2}$   & $8.8\times 10^{-3}$ &   $8.9\times 10^{-3}$\\
$4\times 10^6$& $ z$       & 700           &      0.5     & 0.05  &  $2.2\times 10^{-1}$  &$2.1\times10^{-2}$   & $6.8\times 10^{-3}$ &   $9.4\times 10^{-3}$\\
$4\times 10^6$& $ z$       & 600           &      1.0     & 0.05  &  $2.2\times 10^{-1}$  &$1.0\times10^{-2}$   & $1.9\times 10^{-2}$ &   $8.4\times 10^{-3}$\\
$4\times 10^6$& $ z$       & 600           &      2.0     & 0.05  &  $2.2\times 10^{-1}$  &$8.9\times10^{-3}$   & $1.3\times 10^{-2}$ &   $7.2\times 10^{-3}$\\
$4\times 10^6$& $ z$       & 600           &      4.0     & 0.05  &  $2.2\times 10^{-1}$  &$4.2\times10^{-3}$   & $9.6\times 10^{-3}$ &   $4.0\times 10^{-3}$\\
$4\times 10^6$& $ z$       & 300           &      8.0     & 0.05  &  $2.2\times 10^{-1}$  &$1.3\times10^{-3}$   & $5.8\times 10^{-3}$ &   $2.9\times 10^{-3}$\\
$4\times 10^6$& $ z$       & 300           &      16.0   & 0.05  &  $2.1\times 10^{-1}$  &$-1.7\times10^{-4}$   & $2.4\times 10^{-3}$ &   $6.7\times 10^{-4}$\\
$4\times 10^6$& $ z$       & 300           &      32.0   & 0.05  &  $2.2\times 10^{-1}$  &$-8.0\times10^{-5}$   & $1.3\times 10^{-3}$ &   $9.0\times 10^{-5}$\\
$4\times 10^6$& $ z$       & 300           &      64.0   & 0.05  &  $2.2\times 10^{-1}$  &$-4.0\times10^{-6}$   & $6.2\times 10^{-4}$ &   $6.5\times 10^{-6}$\\
\hline
$1\times 10^5$& $ z$       & 3600          &      64.0   & 0.1  &  $2.6\times 10^{-1}$  &$-5.7\times10^{-6}$   & $7.0\times 10^{-4}$ &   $3.9\times 10^{-7}$\\
\end{tabular}
\caption{Parameters and results of the simulations discussed in this paper.  The error in $\nu_\rmt$ is estimated based on the rms turbulent noise in the Fourier transform at the oscillation frequency. $K$ is the mean kinetic energy density of the fluctuations.}
\label{tab:results}
\end{table*}

\begin{itemize}
\item At sufficiently high frequency, we find
  $\mathrm{Re}(\nu_\rmt)<0$ in every case. Note, however, that noise
  dominates many high-frequency measurements of this quantity. To
  substantiate the high-frequency behaviour, we have performed a
  simulation at lower Rayleigh number ($Ra=10^5$) for more than 3000
  turnover times in order to reduce the effect of turbulent noise.
  This simulation exhibits clearly $\mathrm{Re}(\nu_\rmt)<0$ at
  $\Omega=64$, indicating that the trend observed in the other
  configurations is real.
\item $\mathrm{Im}(\nu_\rmt)>0$. This conclusion is rather strong
  since the noise level is smaller than the measured value of
  $\mathrm{Im}(\nu_\rmt)$ in most cases.  It means that the effective
  elasticity is positive.
\item Larger values of $|\nu_\rmt|$ are found when the stratification
  is in the $x$ direction (shearwise convection). This suggests that
  the full turbulent viscosity tensor is anisotropic.
\item Changing $S$ does not change significantly the measured
  turbulent viscosity, suggesting that the numerical simulations are
  in the linear regime assumed in writing down equation~(\ref{nut1}).
  However, the results obtained with the smaller value of $S$ are
  subject to significant uncertainty.
\item The asymptotic behaviour in the high-frequency limit is
  $\mathrm{Re}(\nu_\rmt)\propto\omega^{-2}$ and
  $\mathrm{Im}(\nu_\rmt)\propto\omega^{-1}$.
\end{itemize}

We have also fitted the simple closure model (\ref{eq:num_closure}) to
the numerical results for shearwise and spanwise convection.  The best
fits are shown as green curves in Figs~\ref{Fig:shearwise}
and~\ref{Fig:spanwise}, and the coefficients we obtained are shown in
Table~\ref{tab:fits}.  The major viscoelastic component has an
effective elastic modulus $c_1$ that is comparable to, but somewhat
less than, the mean kinetic energy density $K$
(cf.~eq.~\ref{eq:iso_elastic}, which suggests a ratio of $4/15$ in the
isotropic case) and a relaxation rate comparable to the nominal
convective turnover rate $|N|$, which is $1$ in our system of units.
Note that the contribution to $\mathrm{Re}(\nu_\rmt)$ from the first
component is $(c_1/\gamma_1)[1+(\omega/\gamma_1)^2]^{-1}$, which is
compatible with the form of the frequency-dependent viscosity used by
\citet{2007ApJ...661.1180O}.  The suggested presence of a second
viscoelastic component with a negative $c_2$ and a faster relaxation
indicates that, at high frequency, the real part of the viscosity
changes sign, as is observed in the numerical results.  With more data
we might be led to introduce a broader spectrum of viscoelastic
components.
\begin{table}
\begin{tabular}{ccc}
\hline
coefficient&shearwise convection&spanwise convection\\
\hline
$c_1$            &  $0.0978$  & $0.06$ \\
$\gamma_1$ & $1.07$                       &  $1.62$ \\
$c_2$           & $-0.0286$   & $-0.0178$ \\
$\gamma_2$ & $6.15$                       & $24.25$ \\
\end{tabular}
\caption{Best-fit coefficients for the closure model (\ref{eq:num_closure}).}
\label{tab:fits}
\end{table}

\begin{figure*}
  \centering
  \includegraphics[width=0.45\linewidth]{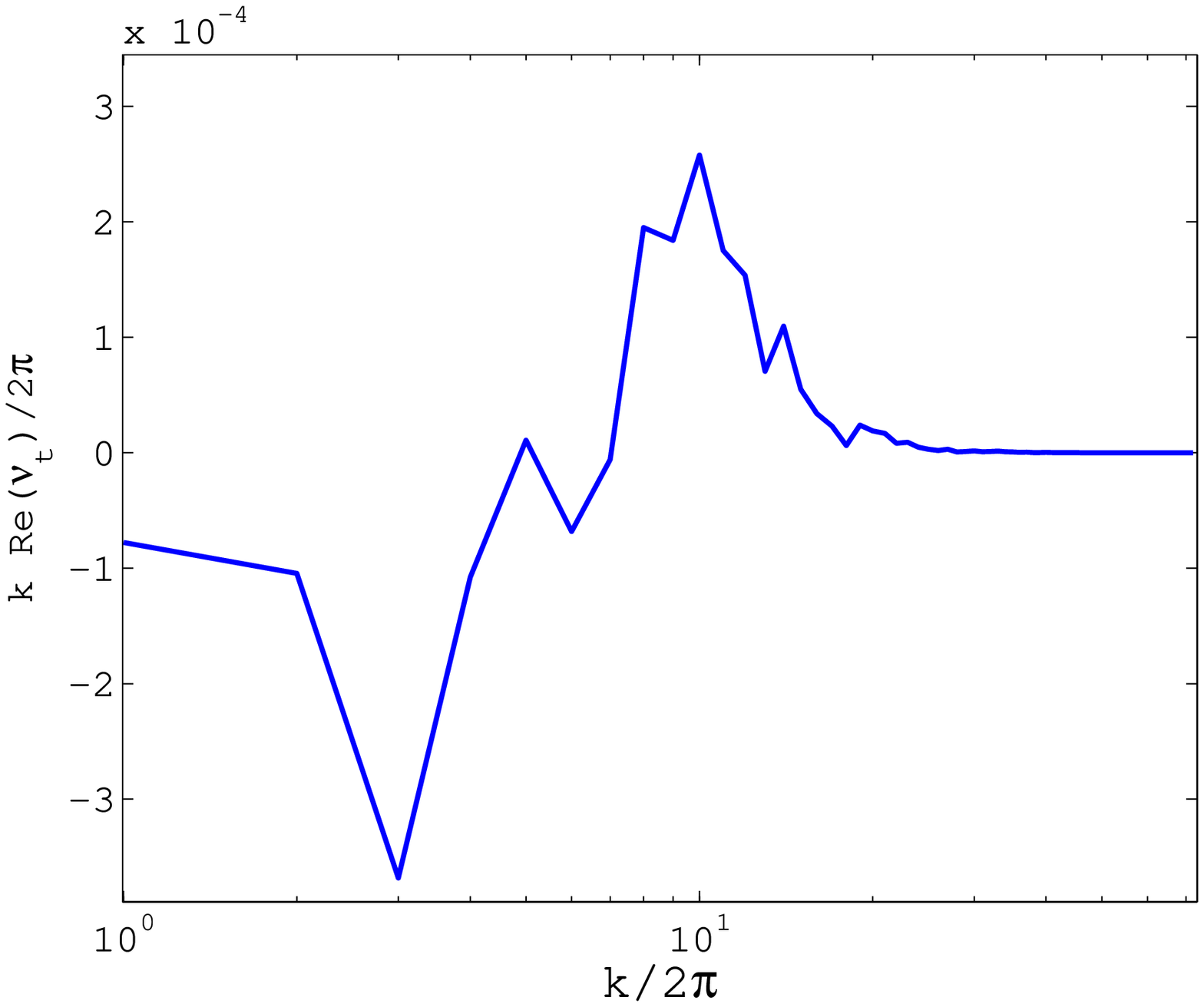}
  \includegraphics[width=0.45\linewidth]{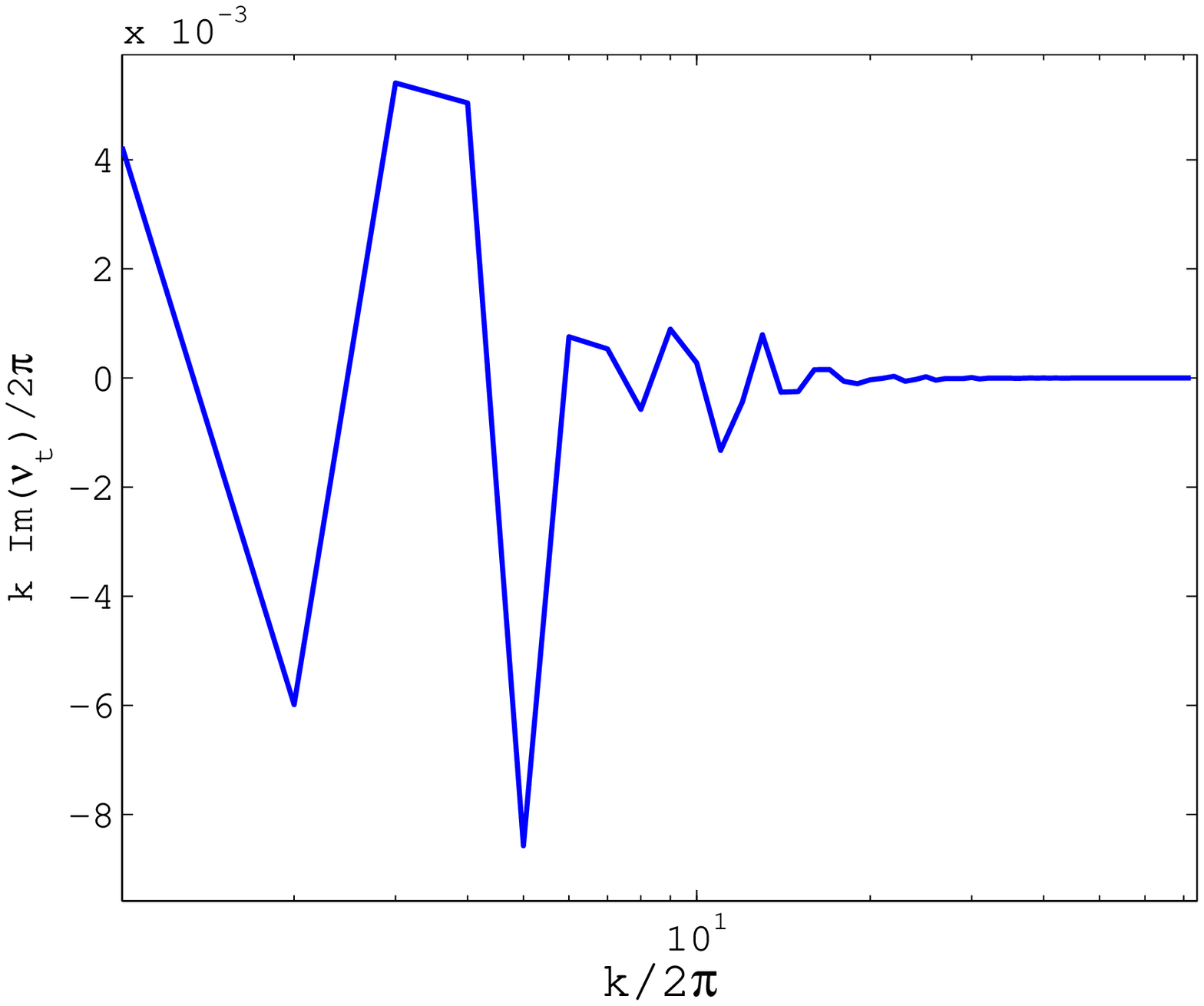}
  \caption{Spatial spectrum of the turbulent viscosity for the
    simulation of spanwise convection with $\Omega=16$ and $S=0.1$.
    The real part has two components: a negative contribution
    from large scales and a positive contribution from smaller scale
    ($k/2\pi\sim 10$).}
  \label{Fig:nuspec}
\end{figure*}

In order to identify which spatial scales contribute to the turbulent
viscosity, we have computed the spatial `spectrum' of the turbulent
viscosity.  This is derived from the instantaneous spatial spectrum of
the Reynolds stress,
\begin{equation}
\widehat{R}_{xy}(k,t)=2\,\mathrm{Re}\Big[\overline{\hat{v}_x\hat{v}_y^*}\Big],
\end{equation}
where $\hat{v}_i$ is the 3D instantaneous spatial Fourier transform of
$v_i$ and the overbar denotes a shell-integration procedure in
spectral space, such that $R_{xy}(t)=\int_0^\infty\widehat
R_{xy}(k,t)\,\rmd k$. $\hat{R}_{xy}(k,t)$ is then used in a relation
similar to (\ref{nut2}) which defines the turbulent viscosity
spectrum $\widehat{\nu_\rmt}(k,\omega)$:
\begin{equation}
  \widetilde{\widehat{R}}_{xy}(k,\omega)=\widehat{\nu_\rmt}(k,\omega)S\,\pi\big[\delta(\omega-\Omega)+\delta(\omega+\Omega)\big].
\end{equation}
We have applied this procedure to the simulation with $\Omega=16$,
$S=0.1$ and with stratification in the $z$ direction. The
corresponding spectra are shown in Fig.~\ref{Fig:nuspec}. As can be
seen, much of the turbulent viscosity (both real and imaginary) is due
to the largest scales of the system. However, we also observe a
component with $\mathrm{Re}(\nu_\rmt)>0$ at `small' scales
($k/2\pi\sim 10$), indicating that there may be an important
contribution from scales whose turnover time is of the order of the
tidal frequency. It should be stressed that these spectra are strongly
polluted by turbulent noise, especially at low wavenumbers. Therefore,
this result should be seen as a plausible trend. Longer simulations
having a larger resolution should be used to confirm this finding.

\section{Summary and discussion}

In this paper we have studied the interaction between tides and
convection in astrophysical bodies by analysing the effect of a
homogeneous oscillatory shear on a fluid flow.  {This model can
  be taken to represent the interaction between a large-scale periodic
  tidal deformation and a smaller-scale convective motion.}  We first
considered analytically the limit in which the shear is of low
amplitude and the oscillation period is short compared to the
timescales of the {unperturbed} flow.  In this limit there is a
viscoelastic response and we obtained expressions for the effective
elastic modulus and viscosity coefficient.  The effective viscosity is
inversely proportional to the square of the oscillation frequency,
with a coefficient that can be positive, negative or zero depending on
the properties of the unperturbed flow.  We also carried out direct
numerical simulations of Boussinesq convection in an oscillatory
shearing box and measured the time-dependent Reynolds stress.  The
results indicate that the effective viscosity falls rapidly as the
oscillation frequency is increased, attaining small negative values in
the cases we have examined.

Our methods and findings differ significantly from those of other
authors.  The hypothesis of \citet{1966AnAp...29..489Z} that the
effective viscosity of large eddies is reduced by only a linear factor
at high frequencies is incompatible with our analytical and numerical
results.  We find a quadratic reduction factor at high frequencies,
similarly to \citet{1977Icar...30..301G} and
\citet{1977ApJ...211..934G}, but not necessarily for the same reason.
Our analytical study, which goes beyond that of
\citet{1997ApJ...486..403G}, implies that large eddies generically
provide a quadratically reduced viscosity [which
\citet{1977Icar...30..301G} considered as an upper bound], but with a
coefficient that can be positive, negative or zero.  Our numerical
results also indicate a quadratically reduced viscosity, with a
tendency towards negative values at high frequencies in the cases we
have examined, an effect due to the largest scales in the turbulent
flow.

The appearance of a negative effective viscosity may be alarming.
However, there is no reason in principle why the effective viscosity
of a convective fluid need be positive.  {The consequence of a
  negative effective viscosity would be that tidal evolution proceeds
  in the reverse direction to that usually assumed.  Angular momentum
  would be transferred in an `anti-frictional' direction, from the
  less rapidly rotating component to the more rapidly rotating one.
  For example, if a small satellite orbits a body with a negative
  effective viscosity, and if the larger body spins more slowly than
  the orbit, then the orbit would expand and the larger body would
  spin down.  Energy would be added to the spin-orbit system.
  (Similarly, orbital eccentricity could be caused to increase in
  situations where circularization would usually be expected.) While
  this effect might be called tidal `anti-dissipation', it does not
  contradict the laws of thermodynamics.  Within limits, work can
  indeed be done by the convection on the tidal flow; the energy comes
  from the buoyancy forces that drive the convection, and ultimately
  from nuclear or gravitational energy.  Since we are considering a
  regime in which the tidal strain is small and the tidal frequency is
  high (implying that the negative effective viscosity is much smaller
  in magnitude than the values estimated from mixing-length theory),
  only a very small fraction of the energy budget of the star would be
  diverted into the spin-orbit system.  The consequences could still
  be important, because the nuclear energy content of the Sun, for
  example, is about five orders of magnitude larger than the orbital
  energy of (say) a solar-type binary star with an orbital period of
  ten days.}

A well known example of a related phenomenon is the negative effective
viscosity of convection at low or moderate Rayleigh number in an
accretion disc \citep[][and references therein]{2010MNRAS.404L..64L},
which would lead to anti-frictional angular momentum transport and
anti-diffusion of the surface density of the disc.  This example also
serves as a reminder that the Coriolis force present in a rotating
system can affect the transport of (angular) momentum; it would
therefore be of interest to include rotation in the calculations
presented in this paper.  {Anti-frictional processes are also well
known in the context of the Earth's atmosphere \citep[e.g.][]{2000McI}.}

{There are several reasons why this reversed tidal evolution
  may not, in fact, be found in nature.  First, it may be that when a
  more realistic tidal velocity field is considered and the actual
  anisotropies of the convection and the Coriolis force are taken into
  account, the net effective viscosity turns out to be positive.
  Second, in a solar-type star where the convective timescale becomes
  short near the surface, the net effective viscosity may be dominated
  by these more rapid convective elements and work out to be positive.
  Third, it may be that tidal dissipation is dominated by internal
  waves and that convection makes a relatively small contribution.
  Fourth, the negative effective viscosities that we find might be due
  to explicit (molecular) viscous effects which are totally negligible
  in astrophysical objects. Indeed, we have not found a negative
  turbulent viscosity that exceeds the explicit (molecular) viscosity
  in magnitude.  Nevertheless, it is of interest to note the
  theoretical possibility of reversed tidal evolution and to be alert
  to any observational evidence of such an effect.}

Penev and collaborators have attempted to address the issue of the
reduction of the effective viscosity at high frequencies through
simulations of convection in a deep layer.  While their results
suggest only a linear reduction factor {(and a positive
  numerical value)}, there are many differences between their work and
ours. Our numerical study involves a local model of convection, which
is in some ways more limited but is also more controlled because a
uniform unstable stratification is imposed.  We measure the effective
viscosity directly rather than relying on a perturbative calculation
or an indirect measurement.  \citet{2009ApJ...704..930P} do not find
good agreement between different methods, and their results depend on
the shear amplitude.  Our work covers a wider range of oscillation
frequencies and quantifies the uncertainty due to turbulent noise.  It
should also be pointed out that our work uses a very small forcing
compared to Penev and collaborators. In particular, the rms velocities
with and without forcing are identical in all our runs, whereas
\citet{2009ApJ...704..930P} have forced velocities of the order of the
rms velocity in their case of weak forcing, and even 5 times larger in
the case of strong forcing. We believe that such strong forcing could
significantly affect the measurement of turbulent viscosity.  Several
tests that we have performed indicate that doubling the forcing
amplitude (to $S=0.2$) is already enough to modify our results.

In our opinion more work is required to extend both analytical and
numerical calculations of this type before the frequency-dependent
effective viscosity of a realistic stellar or planetary convective
zone can be reliably estimated, and the efficiency of tidal
dissipation in astrophysical bodies can be better understood.

\section*{Acknowledgments}

This research was supported by STFC.

\appendix

\newpage

\onecolumn

\section{Composition of an arbitrary traceless velocity gradient tensor from simple shears}
\label{s:proof}

{We consider an arbitrary incompressible velocity gradient
  tensor, being a $3\times3$ matrix with vanishing trace.  Each of the
  off-diagonal elements represents a simple shear, and therefore the
  off-diagonal part of the matrix is a linear combination of simple
  shears.  To compose the diagonal elements, we first consider the
  matrix
\begin{equation}
  \left(\begin{matrix}0&1&0\\1&0&0\\0&0&0\end{matrix}\right),
\end{equation}
which is a linear combination of simple shears.  By a rotation through $\pi/4$ about the $z$ axis, we bring this matrix into the diagonal form
\begin{equation}
  \left(\begin{matrix}1&0&0\\0&-1&0\\0&0&0\end{matrix}\right).
\end{equation}
In a similar way the matrix
\begin{equation}
  \left(\begin{matrix}1&0&0\\0&0&0\\0&0&-1\end{matrix}\right)
\end{equation}
can be constructed from simple shears.  Using a linear combination of
these two diagonal matrices, the three diagonal elements of the
arbitrary velocity gradient tensor can be composed, subject to the
constraint that the sum of the diagonal elements vanishes.}

\section{Vanishing of certain triple velocity correlations for
  statistically isotropic flows}
\label{s:triple}

Let $\bv(\bx)$ be a solenoidal vector field in $d\ge2$ dimensions
satisfying periodic boundary conditions, and let $\langle\cdot\rangle$
be a spatial average over a periodic cell.  (We omit primes on $\bx$
and $\bnabla$ in this Appendix.)  We assume that $\bv(\bx)$ has zero
mean and is statistically isotropic so that any averages of tensor
products of $\bv$ and its derivatives are isotropic
tensors.\footnote{It is debatable whether this assumption is
  compatible with the periodic boundary conditions.  While the cubic
  symmetry imposed by the boundary conditions of a periodic cube are
  compatible with isotropy for tensors of second rank, this is not
  true for higher ranks.  One could argue that the flow can be nearly
  statistically isotropic if it is dominated by scales smaller than
  the box.} The inverse Laplacian operator $\Delta^{-1}$ is well
defined and self-adjoint for periodic fields of zero mean.

We first consider the tensor
\begin{equation}
  D_{abcd}=\langle v_av_b\p_cv_d\rangle.
\end{equation}
As an isotropic tensor, this must be a linear combination of products
of Kronecker deltas:
\begin{equation}
  D_{abcd}=t_1\delta_{ab}\delta_{cd}+t_2\delta_{ac}\delta_{bd}+t_3\delta_{ad}\delta_{bc}.
\end{equation}
The symmetry and contraction conditions $D_{abcd}=D_{bacd}$ and $D_{abcc}=0$
imply $t_2=t_3$ and $dt_1+t_2+t_3=0$.  Thus
\begin{equation}
  D_{abcd}=t_4[d(\delta_{ac}\delta_{bd}+\delta_{ad}\delta_{bc})-2\delta_{ab}\delta_{cd}].
\end{equation}
However, we also have
\begin{equation}
  D_{abba}=\langle v_av_b\p_bv_a\rangle=\langle\p_b({\textstyle\f{1}{2}}v_av_av_b)\rangle=0=t_4d(d-1)(d+2),
\end{equation}
and so
\begin{equation}
  D_{abcd}=0.
\end{equation}

We next consider
\begin{equation}
  D_{abcdef}=\langle v_av_b\p_c\p_d\p_e\Delta^{-1}v_f\rangle.
\end{equation}
Isotropy and symmetry imply
\begin{equation}
\begin{split}
  &D_{abcdef}=t_5(\delta_{ab}\delta_{cd}\delta_{ef}+\delta_{ab}\delta_{ce}\delta_{df}+\delta_{ab}\delta_{cf}\delta_{de})\nonumber\\
  &\qquad+t_6(\delta_{ac}\delta_{bd}\delta_{ef}+\delta_{ac}\delta_{be}\delta_{df}+\delta_{ad}\delta_{bc}\delta_{ef}+\delta_{ad}\delta_{be}\delta_{cf}+\delta_{ae}\delta_{bc}\delta_{df}+\delta_{ae}\delta_{bd}\delta_{cf})\nonumber\\
  &\qquad+t_7(\delta_{ac}\delta_{bf}\delta_{de}+\delta_{ad}\delta_{bf}\delta_{ce}+\delta_{ae}\delta_{bf}\delta_{cd}+\delta_{af}\delta_{bc}\delta_{de}+\delta_{af}\delta_{bd}\delta_{ce}+\delta_{af}\delta_{be}\delta_{cd}).
\end{split}
\end{equation}
(There are three types of term here: those in which the two undifferentiated $v$s are paired, those in which each such $v$ is paired with a $\p$, and those in which one of them is paired with the differentiated $v$.  Symmetry demands that the coefficients of terms of the same type are equal.)

We require the contraction $D_{abcdee}$ (and those related by
symmetry) to vanish.  Thus
\begin{equation}
  (d+2)t_5\delta_{ab}\delta_{cd}+t_6(d+2)(\delta_{ac}\delta_{bd}+\delta_{ad}\delta_{bc})+2t_7(\delta_{ab}\delta_{cd}+\delta_{ac}\delta_{bd}+\delta_{ad}\delta_{bc})=0,
\end{equation}
which implies $t_5=t_6$ and $(d+2)t_5=-2t_7$, and so
\begin{equation}
\begin{split}
  &D_{abcdef}=t_8[(d+2)(\delta_{ac}\delta_{bf}\delta_{de}+\delta_{ad}\delta_{bf}\delta_{ce}+\delta
_{ae}\delta_{bf}\delta_{cd}+\delta_{af}\delta_{bc}\delta_{de}+\delta_{af}\delta_{bd}\delta_{ce}+\delta_{af}\delta_{be}\delta_{cd})\nonumber\\
  &\qquad-2(\delta_{ab}\delta_{cd}\delta_{ef}+\delta_{ab}\delta_{ce}\delta_{df}+\delta_{ab}\delta_{cf}\delta_{de}+\delta_{ac}\delta_{bd}\delta_{ef}+\delta_{ac}\delta_{be}\delta_{df}+\delta_{ad}\delta_{bc}\delta_{ef}+\delta_{ad}\delta_{be}\delta_{cf}+\delta_{ae}\delta_{bc}\delta_{df}+\delta_{ae}\delta_{bd}\delta_{cf})].
\end{split}
\end{equation}

The contraction $D_{abcddf}$ (and those related by symmetry) produces
a Laplacian, so $D_{abcddf}=D_{abcf}$, which we have already shown to
vanish.  Thus
\begin{equation}
  (d+4)t_8[d(\delta_{ac}\delta_{bf}+\delta_{af}\delta_{bc})-2\delta_{ab}\delta_{cf}]=0,
\end{equation}
which implies $t_8=0$ and so
\begin{equation}
  D_{abcdef}=0.
\end{equation}

The last tensor to consider is
\begin{equation}
  D_{abcdefgh}=\langle v_av_b\p_c\p_d\p_e\p_f\p_g\Delta^{-2}v_h\rangle.
\end{equation}
We give an abbreviated argument.  Isotropy and symmetry imply
\begin{equation}
\begin{split}
  &D_{abcdefgh}=t_9(\delta_{ab}\delta_{cd}\delta_{ef}\delta_{gh}+\delta_{ab}\delta_{cd}\delta_{eg}\delta_{fh}+\cdots)+t_{10}(\delta_{ac}\delta_{bd}\delta_{ef}\delta_{gh}+\delta_{ac}\delta_{bd}\delta_{eh}\delta_{fg}+\cdots)\nonumber\\
  &\qquad+t_{11}(\delta_{ah}\delta_{bc}\delta_{df}\delta_{eg}+\delta_{ah}\delta_{bc}\delta_{dg}\delta_{ef}+\cdots).
\end{split}
\end{equation}
Requiring the contraction $D_{abcdefgg}$ to vanish implies $t_9=t_{10}$ and $(d+4)t_9=-t_{11}$.  The contraction $D_{abccefgh}$ should produce $D_{abefgh}$, which we have already shown to vanish.  This is of the form given above, with $t_5=(d+4)t_9+2t_{10}$, $t_6=(d+6)t_{10}$ and $t_7=2t_{10}+(d+4)t_{11}$.  Therefore $t_9=t_{10}=t_{11}=0$ and so $D_{abcdefgh}=0$.

\label{lastpage}

\end{document}